\documentclass[11pt]{article}
\newcommand{\Comment}[1]{{}}
\usepackage{amsfonts,amsthm,amsmath,amssymb,slashed}
\usepackage[textwidth = 480 pt, textheight = 650 pt]{geometry}

\Comment{\usepackage{color}
\definecolor{MyDarkBlue}{rgb}{0.15,0.15,0.45}
\usepackage[linktocpage=true]{hyperref}
\hypersetup{
colorlinks=true,
citecolor=MyDarkBlue,
linkcolor=MyDarkBlue,
urlcolor=MyDarkBlue,
pdfauthor={Asadig Mohammed, 
Jeff Murugan and Horatiu Nastase},
pdftitle={Looking for a Matrix model of ABJM},
pdfsubject={hep-th}
}

\usepackage[numbers,sort&compress]{natbib}
\usepackage{hypernat}
}

\newcommand\ignore[1]{}
\def\one{{\,\hbox{1\kern-.8mm l}}}

\def\Tr{{\rm Tr\, }}

\def\a{\alpha}\def\b{\beta}
\def\g{\gamma}

\def\s{\sigma}

\def\d{\partial}

\newcommand{\Cset}{{\,\,{{{^{_{\pmb{\mid}}}}\kern-.45em{\mathrm C}}}}}

\newcommand{\be}{\begin{equation}}
\newcommand{\bea}{\begin{eqnarray}}

\newcommand{\ee}{\end{equation}}
\newcommand{\eea}{\end{eqnarray}}

\begin{document}

\makeatletter
\@addtoreset{equation}{section}
\makeatother
\renewcommand{\theequation}{\thesection.\arabic{equation}}

\vspace{1.8truecm}

\vspace{15pt}


\centerline{\LARGE \bf Looking for a Matrix model of ABJM   } 
 \vspace{1truecm}
\thispagestyle{empty} \centerline{
{\large \bf Asadig Mohammed${}^{a,}$}\footnote{E-mail address:
\Comment{\href{mailto:asadig@gmail.com}}{\tt asadig@gmail.com}},
{\large \bf Jeff Murugan${}^{a,b,}$}\footnote{E-mail address:
                                 \Comment{\href{mailto:jeffmurugan@gmail.com}}{\tt jeffmurugan@gmail.com}} {\bf and}
    {\large \bf Horatiu Nastase${}^{c,d,}$}\footnote{E-mail address: \Comment{\href{mailto:nastase@ift.unesp.br}}{\tt nastase@ift.unesp.br}} 
    }

\vspace{1cm}



\centerline{{\it ${}^a$ Astrophysics, Cosmology \& Gravity Center,}}
\centerline{{\it Department of Mathematics and Applied Mathematics, 
University of Cape Town}} 
\centerline{{\it Private Bag, Rondebosch 7700,  South Africa}}

\vspace{.4cm}

\centerline{{\it ${}^b$ National Institute for Theoretical Physics (NITheP),}} 
\centerline{{\it Stellenbosch, South Africa}}

\vspace{0.4cm}

\centerline{{\it ${}^c$ Global Edge Institute, Tokyo Institute of Technology,}}
\centerline{{\it Meguro, Tokyo 152-8550, Japan}}

\vspace{.4cm}
\centerline{{\it ${}^d$ 
Instituto de Fisica Teorica, UNESP-Universidade Estadual Paulista,}} \centerline{{\it 
R. Dr. Bento T. Ferraz 271, Bl. II, Sao Paulo 01140-070, SP, Brazil}}

\vspace{1.4truecm}

\thispagestyle{empty}

\centerline{\bf ABSTRACT}

\vspace{.4truecm}

\noindent
Encouraged by the recent construction of fuzzy sphere solutions in the ABJM theory, we re-analyze the latter from the perspective of a Matrix-like model. In particular, we argue that a vortex solution exhibits properties of a supergraviton, while a kink represents a 
2-brane. Other solutions are also consistent with the Matrix-type interpretation. 
We study vortex scattering and compare with graviton scattering in the massive ABJM background, however our results are inconclusive. We speculate on how to extend our results to construct a Matrix theory of ABJM. 

\vspace{.5cm}

\setcounter{page}{0}
\setcounter{tocdepth}{2}

\newpage

\tableofcontents

\setcounter{footnote}{0}

\linespread{1.1}
\parskip 4pt

{}~
{}~

\section{Introduction}

Based on the earlier Bagger, Lambert \& Gustavsson (BLG) construction \cite{Bagger:2006sk,Bagger:2007vi,Bagger:2007jr,Gustavsson:2007vu}
of an M2-brane action in terms of 3-algebras, Aharony, Bergman, Jafferis \& Maldacena (ABJM) \cite{Aharony:2008ug} constructed an ${\cal N}=6$ 
supersymmetric action for the IR of $N$ M2-branes probing a $C^4/Z_k$ singularity, a $U(N)\times U(N)$, level-$k$ Chern-Simons-matter gauge theory whose fields transform in the
bifundamentals. Not only was this the first time that an action for an arbitrary number of M2-branes was written down but it also allowed further insight into the structure of
M-theory, as well as a new and exciting example of the AdS/CFT with its gravity dual being the large $k$ limit of $AdS_4\times S^7/Z_k$, i.e. $AdS_4 \times CP^{3}$. 

But while most of the work on the ABJM model concentrated either on the field theory side, or on the AdS/CFT duality, it is worth bearing in mind that 
the original interest in the multiple M2-brane system was, of course, the potential for a better understanding of M-theory. Indeed, if we are to use the ABJM model 
toward this end, the most natural possibility that comes to mind - since we are, afterall, dealing with a gauge theory of $N\times N$ matrices - is a Matrix theory-type construction. In fact, very much in the spirit of M-theory, it was found in \cite{Nastase:2009ny,Nastase:2009zu}
that the BPS fuzzy funnel solution of pure ABJM, or the vacuum 
fuzzy sphere solution of the massive deformation of ABJM gives rise to a $D4-$brane on $S^2$,  together with the correct small fluctuations action, in the 
classical limit. This $D4-$brane appears as an M5-brane on $S^1/Z_k$ at large $k$, in a similar way to how $D2-$ and $D4-$branes appear in the Matrix theory of Banks {\it et.al.} (BFSS) \cite{Banks:1996vh,Castelino:1997rv,Grosse:1996mz,Kabat:1997im}.  

It seems reasonable therefore to expect that a Matrix theory-type model can be constructed out of the ABJM model. Following the logic of Matrix theory, we need to 
find a classical solution of ABJM corresponding to a spacetime supergraviton. Since such a solution must be localized on the worldvolume, as well as 
in transverse space, it must be a particular type of vortex solution. However, since the ABJM action, unlike its BFSS counterpart, is conformal, we will see 
that a better definition of its Matrix model is given by a maximally supersymmetric deformation of ABJM \cite{Gomis:2008vc}. 
Consequently, after an initial analysis of pure ABJM model, for most of the rest of 
the paper we focus on its maximally supersymmetric deformation. After identifying various spacetime branes as classical solutions of the model,  
we compute supergraviton scattering and compare this with vortex scattering in ABJM. Unfortunately, we find a mismatch between the simplest 
calculations on both sides that manifests in the associated interaction potentials. We will argue that this mismatch arises because a naive application of the BFSS model does not identify the correct calculations on both sides that are supposed to match. We will then speculate on the Matrix theory rules for the correct identification of the two sides.

The paper is organized as follows. In section 2, in the interests of being self-contained, 
we provide a lightning review of BFSS theory, focusing on those elements that carry over to our case, and then in section 
3 we will describe in detail our set-up and what we expect to find. Section 4 is devoted to an analysis of solutions of pure ABJM theory, and we argue that, 
while it is possible to identify a vortex solution with a supergraviton, such a solution is practically unfeasible due in no small part to the infinite energy of the corresponding background D2-brane. Instead in section 5 we will focus on the massive deformation of ABJM and 
identify its known solitonic spectrum (consisting of brane-filling, kink and vortex types) with branes in spacetime. In the process we identify the background in which the M2-branes 
corresponding to the massive ABJM move.
In section 6 we calculate the scattering of supergravitons in this massive background, and in section 7 the corresponding scattering of vortices in massive ABJM. Finally, after noting a mismatch of the calculations of sections 6 and 7, in section 8 we speculate on the possible definition of the sought-for ABJM Matrix theory. 

\section{A short review of the BFSS matrix theory}

The BFSS Matrix theory of \cite{Banks:1996vh}
is based on a discrete light cone quantization (DLCQ) of M-theory with a compact circle. In the heuristic derivation 
given by Sen and Seiberg \cite{Sen:1997we,Seiberg:1997ad}, this light-like compactification of the given M-theory is related to a space-like 
compactification of a different M-theory in a decoupling limit, in which the only thing that remains is a decoupled theory of $N$ D0-branes. Each D0-brane in this description corresponds to a single unit of momentum in the compact (11th) M-theory direction. Its action is given by
\be
  S=\int dt\,\, \Tr\left[\frac{1}{2R}D_t \tilde X^i D_t \tilde X^i-\frac{RM_P^6}{4}
  \left[\tilde X_i,  
  \tilde X_j\right]^2-\theta^TD_t\theta -RM_P^2 \theta^T\gamma_i
  \left[\theta,\tilde X^i\right]\right]\,,
  \label{BFSSaction}
\ee
where the $\tilde X^i=X^i/g_s^{1/3}$ are nine scalars corresponding to the nine transverse directions, $D_t=\d_t+iA$, and $A$ is a 0+1 dimensional $U(N)$ gauge field. In addition $R$ is
the compactification radius, $M_{P}$, the 11-dimensional Planck mass and the $\theta$ are fermionic superpartners of the $\tilde{X}^{i}$ that transform as spinors under the $SO(9)$ group of transverse rotations.
 
The simplest classical solution of the BFSS Lagrangian 
\be
 \tilde X^i(t)=\left(x_0^i+v^it\right)\mathbb{I}_{M\times M}\,, 
 \label{BFSSgrav}
\ee
can be understood as $M\leq N$ D0-branes located at $\vec{x}_0$ and 
moving with velocity $\vec{v}$. In spacetime, this corresponds to a pointlike object with $M$ units of momentum along the 11th direction\footnote{Or, equivalently, carrying 
D0-brane charge $M$.} and is interpreted as a supergraviton. Other classical solutions correspond to other $D-$branes in spacetime - $D2-$branes and $D4-$branes - with different possible geometries.

Arguably, the calculation that received most attention in Matrix theory, ostensibly providing the first real test of the BFSS conjecture, is the matching of the interaction 
potential of two spacetime supergravitons with the corresponding interaction potential for two corresponding objects of the form (\ref{BFSSgrav}) in Matrix theory. The calculation of the supergraviton interaction potential, as described in \cite{Becker:1997cp}, 
is based on the observation of 't Hooft \cite{'tHooft:1987rb} that 
the tree level Rutherford scattering interaction potential in gravity (mediated by single graviton exchange) can be calculated by scattering two gravitational
shockwaves. One of these is described by an Aichelburg-Sexl shockwave - which serves as a heavy {\it source graviton}, while the second is a plane wave probing it. Note that the 't Hooft calculation was in a flat space background, but this procedure was also applied successfully to the curved space case \cite{Kang:2004yk,Kang:2004jd}.

The first step in determining the interaction potential in the more general case, then, is 
to calculate the shockwave in the given spacetime background. This plays the role of the graviton wavefunction and corresponds to adding an $h_{--}(dx^-)^2$ term to the 
background metric. These `pp' shockwaves have the remarkable property (not shared by many solutions in the highly nonlinear field equations of general relativity) that the linearized solution is exact, i.e. $h_{--}$ is the solution of the Poisson equation
\be
\Delta^\perp_{bgr}(\vec{x}) h_{--}(\vec{x})=Q\delta^\perp(\vec{x})\,,
\ee
where the source $Q$ depends on the momentum of the wave in the 11th dimension,  $p_{11} = N_s/R$ and $M_P$. In the BFSS case, there are nine transverse directions (and two parallel directions -  time and the 11th dimension, in which the wave propagates), so, by dimensional analysis, the solution must scale like $1/r^7$. More precisely $\displaystyle h_{--} = \frac{15\pi N_{s}}{RM_{P}^{9}r^{7}}\delta(x^{-})$, and the source graviton is in a state of definite $p_{-}$ so that averaging over $x^{-}\in(0,2\pi R)$ gives
\be
  h_{--}=\frac{15 N_s}{2R^2 M_P^9}\frac{1}{r^7}\,.
\ee
To compute the interaction potential we then scatter a probe scalar (plane wave) off the shockwave background. Practically speaking, this requires
considering the $h_{--}$ part of the metric as an interaction and using standard quantum mechanical perturbation theory. Consequently, the free wave equation $\Box \phi=0$ becomes
\be
  \left[\frac{\d^2}{\d\vec{x}^2}-\frac{\d^2}{\d x_0^2}+m^2\right]\phi=\left[-\frac{h_{--}}  
  {2}(\d_0-m)^2\right]\phi\equiv V \phi\,
\ee
where $\phi =\bar \phi e^{ip x_{11}}$ and $p=im$. At this stage, a few points deserve some elaboration. 
\begin{itemize}
  \item
  Momentum in the $11^{th}$ dimension acts like a mass from the 10-dimensional point of view.    
  \item
  Even though gravitons move at the speed of light, we call a graviton moving entirely in the 
  $11^{th}$ dimension a $v=0$ solution. This will be an $h_{--}$ gravitational perturbation. 
  \item
  The probe graviton has nonvanishing $v$ (i.e., a relative velocity between the probe 
  and the source), and so will also propagate ever so slightly in the other ten dimensions. 
  \item
  For both gravitons, however, $x^-$ plays the role of time which is as it should be seeing as 
  how we are in a DLCQ description. 
\end{itemize}
With this prescription (and the usual relativistic normalization), the probe graviton wavefunction 
\be
\phi(x)=\frac{1}{(2\pi)^{9/2}}\frac{1}{\sqrt{2E}}e^{i\vec{p}\vec{x}-iEx_0}\,,
\ee
where $E^2=\vec{p}^2+m^2$, $E =\frac{m}{\sqrt{1-v^2}}$ and $\vec{p}=\frac{m\vec{v}}{\sqrt{1-v^2}}$. The 1-loop interaction potential in momentum space is computed through 
the S-matrix
\be
  2\pi \delta(E_{in}-E_{out})V_{int}^{(1)}=S^{(1)}=\int d^{10}y\,\, \phi_{out}(y)^*V(y)
  \phi_{in} (y).
  \nonumber
\ee
On substituting for the graviton wavefunction and the expression for $V(y)$ above,
\bea
S^{(1)}=\frac{i(\tilde{E}_{in}-m)^2}{(2\pi)^92\sqrt{E_{in}E_{out}}}\int dx_0\,\, e^{i(E_{out}-E_{in})x_0}
\int d^9 x\,\, e^{i(\vec{p}_{in}-\vec{p}_{out})\vec{x}}h_{--}(\vec{x})\,,
\eea
so that after integrating, the interaction potential can be read off as 
\bea
V^{(1)}_{int}(\vec{x})=h_{--}\frac{(\tilde{E}-m)^2}{2\tilde{E}}\simeq m\frac{v^4}{8}h_{--}+{\cal O}(v^6)\,.
\label{bfsspot}
\eea
Finally, using $m=N_p/R$ gives the full one-loop result 
\be
V_{int}(\vec{x})\sim N_sN_p\frac{v^4}{r^7}\,.\label{BFSSintpotential}
\ee
It is this interaction potential - in particular the coefficient of the $v^{4}/r^{7}$ term - that is then matched to the one-loop Matrix-theory of BFSS 
\cite{Becker:1997cp}. To compute the interaction potential of two supergravitons 
\be
X^i(t)=(x_1^i+v_1^it)\mathbb{I}_{N_1\times N_1}+(x_2^i+v_2^it)\mathbb{I}_{N_2\times N_2},
\ee
separated by $r=|\vec{x}_1-\vec{x}_2|$ and with relative velocity $v=|\vec{v}_1-\vec{v}_2|$ in Matrix theory, we note that since the two diagonal blocks are {\it non-overlapping} these can be interpreted as two {\it distinct} supergravitons with correspondingly different extent in the transverse directions. If we choose $v_1^i=-v^i$ and $v_2^i=+v^i$, we get that
\be
H=\frac{1}{2R}\Tr |\dot{X}^i|^2=\frac{(N_1+N_2)}{R}\frac{v^2}{2}=\frac{m_1v^2}{2}+\frac{m_2v^2}{2}=E_1+E_2\,,
\ee
from the Matrix action. This is the free nonrelativistic energy of the supergravitons. To find the interaction potential, we must go to one-loop, in which case one takes the one-loop fluctuation determinant around 
\be
  X^i(t)=(b^i+v^it)\mathbb{I}_{N\times N}\,,
\ee
where now $b^i=x_1^i-x_2^i$ and $v^i=v_1^i-v_2^i$ are the relative positions and velocities respectively. The calculation of one-loop determinants amounts to just the zero point fluctuation $\sum_n \omega_n/2$. Actually, this is the same calculation that one does to compute the quantum mass of solitons like, for example, a kink. There one calculates this sum in the background of the kink solution
(see, e.g., \cite{Nastase:1998sy}). The result is in perfect agreement with (\ref{BFSSintpotential}). 

To summarize then, the leading order calculation of the interaction potential on the Matrix theory side matches the leading order result on the gravity side, even though 
the former is at {\it one-loop}, while the latter is {\it classical}. At this point, it is only natural to ask:

{\it Is there an analogous computation of interaction potentials that can be performed in ABJM?}

\section{Set-up and expectations}

We saw that the BFSS model - a (0+1)-dimensional $U(N)$ Matrix model on the worldvolume of $N$ D0-branes - describes M-theory in discrete light cone 
quantization. However, since the $D0-$branes are momentum modes on the compact 11th dimension, this description of M-theory is not a fundamental one. Instead,
as shown by Sen and Seiberg in \cite{Sen:1997we,Seiberg:1997ad}, 
it appears because of the equivalence of the original M-theory with a decoupled theory of D0-branes living in another 
M-theory. Any fundamental description of M-theory must involve M2-branes instead, but we don't know how to formulate it.
In the large $N$ limit classical $D-$branes appear as solutions of the classical BFSS theory. For example, D2-brane solutions in \cite{Banks:1996vh} were found, wrapping a fuzzy torus defined through, 
\bea
  &&X^8=R_8 P;\;\;\;
  X^9=R_9 Q\cr
  &&P=\sqrt{N}p;\;\;\;
  Q=\sqrt{N}q\cr
  &&[Q,P]=2\pi i
\eea
where $U=e^{ip}$ and $V=e^{iq}$ are the ``clock" and ``shift" operators of the fuzzy torus 
satisfying 
\bea
&&UV=e^{\frac{2\pi i}{N}}VU\cr
&& U^N=V^N=1
\eea
Another instructive example is Matrix theory in a pp-wave background ({\it i.e.} the BMN Matrix model \cite{Berenstein:2002jq}). Here, $2-$branes wrapping a fuzzy $S^2$ are also a solution. There are two ways to think about this: either as another example of the same BFSS construction as BFSS, only in a different spacetime background, or as a massive deformation of the BFSS model since the pp-wave corresponds to a mass deformation on the brane worldvolume. 
The presence of the mass deformation also serves to better define the Matrix theory as it makes states discrete instead of continous. If Matrix theory is to correctly describe M-theory (and its dimensional reduction to type IIA string theory) then it should be able to describe {\it all} $D-$branes in the theory and not just $D2-$branes. 
For example, a $D4-$brane wrapping an $S^4$ was found 
in \cite{Castelino:1997rv}, following the earlier works of \cite{Grosse:1996mz,Kabat:1997im}, but the solution is not without several unresolved subtleties.
In general, finding the complete spectrum of $D-$branes from Matrix theory remains a very difficult problem.
The $D2-$ and $D4-$branes already found are reductions to ten dimensions of $M2-$ and 
$M5-$branes, and while they are a minimum necessary for the spectrum of M-theory, they are by no means sufficient. 
Indeed, we would also need to find a $D6-$brane, coming from an eleven dimensional KK monopole, and a $D8-$brane\footnote{The latter would appear in the massive type IIA 
string theory, for which the Matrix theory was constructed in \cite{Lowe:2003qy}. Its properties remain largely unexplored.}.

Fortunately, the recent construction by Aharony {\it et.al.} \cite{Aharony:2008ug} 
of an IR action for $N$ $M2-$branes at a $C^4/Z_k$ singularity offers some much needed hope. Since this ABJM model is also a theory of $N\times N$ matrices\footnote{Albeit one with a $U(N)\times U(N)$ gauge group and
bifundamental degrees of freedom.}, it is natural to ask whether one can construct a new kind of Matrix theory from the ABJM model, perhaps one whose formulation is more fundamental instead of the somewhat derived one of BFSS. In support of this idea, it was shown in 
\cite{Nastase:2009ny,Nastase:2009zu} that $D4-$branes wrapping a fuzzy $S^2$ appear as 
solutions of (a massive deformation of) the ABJM model, in much the same way as in the BFSS Matrix theory. In as much as the BFSS Matrix theory is related to the maximally supersymmetric BMN Matrix model by a massive deformation of the former, the pure ABJM model can be mass-deformed to yeild a maximally supersymmetric massive ABJM model \cite{Gomis:2008be}. In this case the fuzzy funnel solution of the pure theory stabilizes to a fuzzy sphere in the massive one. This hints then, that the massive deformation gives a 
better definition of a proposed Matrix model. Nevertheless we will first begin with an analysis of the pure ABJM model and then show how we are driven to its massive deformation. Our strategy will be as follows:
\begin{itemize}
  \item
  The presence of the D4 wrapping a fuzzy $S^2$ solution suggests that we start with a 
  similar set-up to BFSS. 
  \item
  The first issue to be checked is whether there is a classical solution corresponding to 
  a supergraviton, for which we can then compute scattering. 
  \item
  Thereafter, we will test whether other D-branes appear as classical solutions\footnote{And
  reserve the right to not be surprised if, as for BFSS, it turns out to be difficult to find 
  all of them.}.
\end{itemize}
We immediately notice a difference from BFSS; namely that the $N$ $M2-$branes of ABJM have two spatial worldvolume directions while a supergraviton 
needs to be pointlike in both the transverse and parallel space directions. It also carries
$D0-$brane charge $M\leq N$ so we must look for a vortex-type solution of ABJM, pointlike in the transverse space, and with some $D0-$brane charge $M$. Once we find such objects, we can then try to scatter them, and match against scattering of supergravitons.

Even having identified which objects to scatter, the issue of matching superpotentials is still a little murky since there is no {\it a priori} guide which term should correspond to 
which. This is not unique to our ABJM computations either. Indeed, in the BFSS model the leading supergraviton scattering is a tree-level (classical) interaction, while on the Matrix side the leading term is a one-loop interaction with the classical interaction vanishing.
Moreover, on the gravity side the tree-level interactions yield an infinite series that should match various higher loop correction terms in the Matrix side but the precise matching cannot be guessed before doing at least a computation of the $N_s, N_p$ and $M_P$ dependences. 

In our case the increased computational complexity means that even the $N_s, N_p$ dependence is hard to obtain in the ABJM side while the $M_P$ dependence cannot be guessed  before computing the $r$ dependence, as we discuss below.
Our goal therefore will be to first compute and match the simplest, leading terms on both sides. On the gravity side, this will be a similar tree-level term as in the BFSS case, but now in the Matrix model, the leading classical interaction (for the vortex scattering) is no longer vanishing.

One other difference between the ABJM and BFSS models is that pure ABJM is a conformal theory while the BFSS Lagrangian, having $M_P^3$ as coupling, is not. In the BFSS interaction potential, the $M_P$ dependence appears from two sources, one being as 
a coupling dependence, and the other as giving the unit of length when translating from BFSS to gravity. The latter is of the same kind as in AdS/CFT where the $l_s$ dependence of gravity calculations also appears by introducing a unit of length in the conformal calculations of the
field theory. Since pure ABJM is also conformal, only the latter type of $M_P$ dependence would be available. So the matching 
is {\it a priori} more constrained than in the BFSS case. Of course, as we said above, the massive ABJM will be found to be more useful for defining a Matrix model. 
There we will see that we have one more parameter - the mass deformation $\mu$ - potentially improving the situation.
We will return to these issues in some detail in section 8 where we describe the systematics 
of matching in BFSS and how they might apply to our case. 

While on the topic of expectations, it is worth asking at this point what other $D-$branes we expect to find as classical solutions of the ABJM model? Certainly, as in BFSS, we would want at least an example of a $D2-$brane and a $D4-$brane, which would mean that $M2-$ and $M5-$branes appear in the theory. These are already present however; for the $D2-$branes there is at least the configuration of the ABJM $2-$branes themselves\footnote{See, for instance \cite{Mukhi:2008ux} for an example of how the worldvolume theory of $N$ $D2-$branes arises.} while for $D4-$branes there is (at least) the solution corresponding to a $D4-$brane on a fuzzy $S^2$. Of course, this is by no means satisfactory, in part because we now have the peculiar fact that directions parallel and transverse to the ABJM worldvolume are different so we would like to see at least one example of a ``shifted" brane. For example, we would 
expect to find {\it kink} solutions having a fuzzy one-dimensional transverse space and a fuzzy three-dimensional transverse space, corresponding to 
$D2-$ and $D4-$branes respectively, each with only one direction parallel while transverse $D2-$ and $D4-$branes with no parallel directions should be described as {\it vortex} solutions with  fuzzy two-dimensional and a fuzzy four-dimensional transverse space. We will seek to find at least one example among these solutions.

\section{Pure ABJM solutions}

Solutions of the pure ABJM theory and its massive deformation appear to be related, usually in a fairly nontrivial way. For example in \cite{Nastase:2009ny,Nastase:2009zu}, it was shown that not only does the maximally supersymmetric fuzzy sphere 
ground state of massive ABJM becomes the $\frac{1}{2}$-BPS fuzzy funnel solution of pure ABJM, but they also enjoy many shared properties, 
like the same unrescaled bosonic action for fluctuations. 
Certainly then, we would not be surprised if the vortex and kink solitons found in both pure and massive ABJM 
\cite{Arai:2008kv,Kim:2009ny,Auzzi:2009es}, turn out to be similarly related. In this section we will discuss the solitons of pure ABJM, 
though for technical reasons we will be forced to switch to massive ABJM in the next section.

To this end, we begin by asking if we can find vortex solutions of {\it pure ABJM theory} that can be interpreted as supergravitons in spacetime? A good place to start answering this question is with the ansatz presented in \cite{Kim:2009ny} for a general class of vortex solutions,
\bea
Y^A&=& v^A \mathbb{I},\;\; {\rm for} \;\; A=2,3,4\,,\nonumber\\
Y^1&=& \sum_{a=1}^{N-1}y_a e^a +y_M E^{-M}\,,\nonumber\\
y_M&=& \frac{G(z)}{\prod_{a=1}^{N-1} y_a}\,,\nonumber\\
{}\\
\d\bar \d \ln |y_a|^2&=&4v\left(\frac{2\pi}{k}\right)^2\sum_{b=1}^{N-1}K_{ab}\left(|y_b|^2-\frac{|G(z)|^2}{|c_b|^2\prod
_{c=1}^{N-1}|y_c|^2}\right)\,,\nonumber\\
F_{12}=\hat{F}_{12}&=&-\frac{s}{2}\left(\frac{2\pi v}{k}\right)^2\left[Y^1,Y^\dagger_1\right]\,,\nonumber\\
D_0 Y^A&=&is\frac{2\pi}{k}v^A\left[Y^1,Y^\dagger_1\right]\,,\nonumber
\label{defineqs}
\eea
where $Y^A$ are the four complex $U(N)$-valued bifundamental scalars of ABJM; $A_\mu$ and $\hat A_\mu,\, (\mu=0,1,2)$ are worldvolume $U(N)\times U(N)$ 
gauge fields and $e^a,h^a$ and $E^{-M}$ are generators satisfying
\be
(e^a)^\dagger=e^{-a},\,\left(E^M\right)^\dagger=E^{-M},\, \left[E^{-M},e^a\right]=0,\,\left[e^a,e^{-a'}\right]
=\delta_{aa'}h^a,\,\left[h^a,e^b\right]=K_{ab}e^b.
\ee
$K_{ab}$ is the Cartan matrix and we have defined
\bea
&&A=\frac{(A_1-iA_2)}{2}\,,\nonumber\\
&&z=x_1+ix_2\,,\nonumber\\
&&D_\mu Y^A=\d_\mu Y^A+iA_\mu Y^A-iY^A\hat A_\mu\,,\\
&&A=\sum_{a=1}^{N-1}A_a h^a\,,\nonumber\\  
&&v^2=\sum_{A=2}^4|v^A|^2\,.\nonumber
\eea
$G(z)$ is an arbitrary holomorphic function, 
$c_b$ is a constant and $x_1,x_2$ are the two space worldvolume directions. 
The vortex solution then obtains by solving the above equations for the scalars and the gauge fields. There are two particularly simple cases:

\begin{enumerate}
  \item
  By taking the trivial value for the holomorphic function, namely $G=0$, we can obtain 
  an approximate solution. 
  Define $K_a=\sum_b K_{ab}$. Then $y_M=0$ and the equation for $y_a$ becomes 
  $\d\bar \d \ln |y_a|^2=A\sum_b K_{ab}|y_b|^2$, with $A$ a constant. Consider moreover the   
  case when $K_a$ is 
  independent of $a$, in which case we can choose a solution with $g=|y_a|^2$
  independent of $a$, so that $\d\bar\d \ln g= AK_a g$. Then near $z=0$, the solution is   
  approximately $\displaystyle |y^a|^2\sim e^{AK_a |z|^2}$ or,
  \bea
     Y^1\sim \sum_{a=1}^{N-1}e^a e^{\frac{1}{2}AK_a|z|^2}
  \eea
  We will not calculate the gauge fields, since it is not clear how to interpret this approximate   
  solution. 

  \item
  The simplest exact solution is obtained for the simplest nontrivial $G$. In this case, taking
  \be
    y^a=\frac{y^M}{c_a}\,,\nonumber
  \ee
  reduces the $y^{a}$ equation to $\d\bar\d \ln \left|y^M\right|^2=0$ and from the $y^M$ equation   
  we get 
  \be
    y_M=\left(G(z)\prod_{a=1}^{N-1} c_a \right)^{1/N}\,,
  \ee
  Then, if we choose the simplest holomorphic function
  $G(z)=z-z_0$, the equation $\d\bar\d \ln \left|y_M\right|^2=0$ is identically satisfied
  and we have a complete solution. Specifically,
  \bea
    Y^1&=&\left(\sum_{a=1}^{N-1}\frac{e^a}{c_a}+e^{-M}\right)\left((z-z_0)\prod_{a=1}^{N-1}     
    c_a \right)^{1/N}\,,\nonumber\\
    {}\\
    Y^A&=& v^A \mathbb{I},\;\; A=2,3,4.\nonumber
  \eea
\end{enumerate}
For this solution can be interpreted as a supergraviton, we have to first argue that it corresponds to a {\it classical pointlike object} in spacetime. This is not too difficult to see. 
First notice that the corresponding coordinate $Y^A$ is fixed and, according to the usual Matrix theory definitions, a VEV proportional to the identity corresponds to a fixed classical coordinate. Further, at the position of
the vortex, $z=z_0$, $Y^1=0$ is also fixed, so that this object is extended only in time and not in any of the parallel or transverse coordinates.

Let's now calculate the gauge fields for this solution. To do so, we need to compute $A_0,\hat A_0, A$ and $\hat A$ (or equivalently, $A_1,A_2,\hat A_1,\hat A_2$).
Since $F_{12}=\hat F_{12}$, we can choose $A=\hat A$.
First, note that $F_{12}=\d_1A_2-\d_2A_1$ implies that $\d \bar A=\frac{1}{4}[\d_1A_1+\d_2A_2+iF_{12}]$.
Consequently, in the Coulomb gauge, $\d_i A_i=\d_1 A_1+\d_2 A_2=0$, 
the magnetic field takes the sussinct form $B=F_{12}=4\d \bar A/i$. Using the properties of the generators, 
$\left[Y^1,Y^\dagger_1\right]=\sum_{a=1}^{N-1}|y_a|^2 h^a$, so that
substituting into (\ref{defineqs}) and integrating, we get
\bea
B&=& -\frac{1}{2}\left(\frac{2\pi v}{k}\right)^2\left(\prod_{b=1}^{N-1}|c_b|^2\right)^{1/N}
\sum_{a=1}^{N-1}\frac{|z-z_0|^{2/N}}{|c_a|^2}h^a\,,\nonumber\\
{}\\
\frac{A_1+iA_2}{2}&\equiv&
A^*=-\left[\frac{Ni}{8(N+1)}\left(\frac{2\pi v}{k}\right)^2\left(\prod_{b=1}^{N-1}|c_b|^2\right)^{1/N}\right]
\sum_{a=1}^{N-1}\frac{(z-z_0)^{\frac{N+1}{N}}(\bar z-\bar z_0)^{1/N}}{|c_a|^2}h^a\,,\nonumber
\eea
On the other hand, to solve for the $A_0$, we substitute the expression for the commutator $[Y^1,Y^\dagger_1]$ into
\be
  D_0Y^A=iA_0Y^A-iY^A\hat A_0=i\frac{2\pi }{k}v^A\left[Y^1,Y^\dagger_1\right]\,,
\ee
to obtain 
\be
  A_0Y^A-Y^A\hat A_0=Y^A\frac{2\pi}{k}\left(\prod_{b=1}^{N-1}|c_b|^2\right)^{1/N}  
  \sum_{a=1}^{N-1}\frac{h^a}{|c_a|^2}
  |z-z_0|^{2/N}\,,
\ee
from which (after choosing $\hat A_0=0$) we can read off  
\be
A_0=\frac{2\pi}{k}\left(\prod_{b=1}^{N-1}|c_b|^2\right)^{1/N}\sum_{a=1}^{N-1}\frac{h^a}{|c_a|^2}
|z-z_0|^{2/N}\,.
\ee
Note that for this vortex solution, the magnetic field $B=F_{12}$ goes to zero at $z=z_0$ (the position of the vortex), but the electric field $F_{0z}=-\d A_0$ diverges at that same location.

This vortex solution looks like it could stand in for the spacetime supergraviton, since it is a pointlike object in spacetime, carrying $D0-$brane charge. This last property is perhaps not obvious. However, as a massive, classical, pointlike spacetime object in ten dimensions this vortex must carry a charge corresponding to its momentum in the eleventh dimension and, as there is no other possible candidate, this must be identified with a $D0-$brane charge. As we don't have a dual description of this solution though, it is not easy to check this assertion explicitly. We will see that in the related case of the vortex of the massive deformation of ABJM that the corresponding object does indeed carry D0-brane charge. 

However, it is also easy to see that the solution has infinite energy, since the energy density of the magnetic field increases away from $z=z_0$. Note also that the solution has the complex coordinate $Y^1\propto (z-z_0)^{1/N}$ which, in the large $N$ limit exhibits a step function-like behaviour (0 at the vortex position and 1 away from it). In other words, away from the vortex, the solution represents a two-dimensional worldvolume growing at infinity, which could be identified with a 2-brane. As similar as they sound, this solution is different from BIonic branes of \cite{Callan:1997kz,Cook:2003rx}, where a single coordinate $X\sim 1/r$ signals a string extending to infinity at $r=0$, as well as the self-intersecting $M2-$brane of \cite{Callan:1997kz,Gibbons:1997xz,Gauntlett:1997ss}, where the two complex coordinates\footnote{One of these is a worldvolume coordinate, $t=x_1+ix_2$ while the other, $s=X^4+iX^5$, is transverse.} 
$s$ and $t$ are related by $s\propto c/t$ ($c$= constant). In each of those cases, a new brane or string ``grows" at the position of the singularity, and represents the spacetime intersection of branes. Charge conservation then implies that the worldvolume flux must flow through a string or brane which must either extend to infinity, as for the BIon and self-intersecting M2, or end on yet another brane \cite{Callan:1997kz,Cook:2003rx}. This latter solution, ending at $r=r_0$ and finite $X(r_0)$, is half of the $D\perp F1\perp \bar{D}$ brane configuration, and provides an example of charge conservation by ending on another brane with $X(r_0)$ finite and $X'(r_0) = \infty$. 

The vortex solution of pure ABJM is clearly an example of the latter. To be precise, since  $Y^1(z_0)=0$ but ${Y^1}'(z_0)=\infty$, the only way to enforce spacetime charge conservation 
would be to have D0-branes at the endpoint. The situation is muddied however by the fact that $Y^1$ grows in the direction of the original 2-brane (as $z\rightarrow \infty$).

In \cite{FIKS}, it was demonstrated that the vortex solutions of pure ABJM described here have the same supersymmetries as self-intersecting $M2$-branes. 
This fact alone likely means that we can interpret the solution away from $z_0$ as a self-intersecting $M2$-brane, with the caveat that $D0$-branes be added to the $z=z_0$ point 
to enforce charge conservation. It seems then that the vortex solution should be thought of as a bound state of self-interesecting 2-branes and $D0$-branes at $z_0$. This interpretation matches nicely with the picture we will find in the massive ABJM case, where the vortex is a bound state of dielectric $D2$-branes blown up into $D4$-branes, and $D0$-branes at the vortex position. The difference of course, is that in this case the 2-branes are infinite in extent, and have correspondingly infinite total energy.

To summarize; {\it even though the vortex solution looks like a supergraviton, it is hard to get actual physics from it}. Essentially, this is because we will still need to have some appropriate regularization that to render the 
energy finite before we can subtract the 2-brane contribution and interpret the remainder as the supergraviton. There is no reason why this cannot be done in principle, we just have not found a simple way to do it. By contrast, it turns out to be much easier to work with the massive ABJM theory where everything is wonderfully finite.  To this end then, in the next section we study
the finite energy solutions of the massive ABJM theory.

\section{Massive ABJM solitons}

Following several earlier works on massive deformations of BLG theories \cite{Bagger:2007vi,Gomis:2008cv,Hosomichi:2008jb}, a maximally supersymmetric (${\cal N}=6$) massive deformation of ABJM was recently proposed in \cite{Gomis:2008vc}. However, it is not clear exactly what the brane interpretation of the deformed theory is. One suggestion is that before the brane backreaction, the background corresponds to the maximally supersymmetric 
type IIB pp-wave. In \cite{Auzzi:2009es},
a different conjecture, for the gravity dual of massive ABJM was proposed, based on a 
$Z_k$ orbifold of the LLM solution corresponding to the massive deformation of 
$N$ $M2$-branes in flat space. 

For our purposes, we will need to understand the massive deformed background in which the $N$ $M2$-branes move ({\it before} back-reaction) {\it i.e.} the analog of the $C^4/Z_k$ orbifold for pure ABJM. Once we have computed the background dual to the deformed ABJM model, we proceed to analyze the various solutions of massive ABJM. From the point of view of the ABJM worldvolume, these may be classified as brane-filling, vortex or kink type, according to their codimension (zero, two and one respectively). We will provide a brane interpretation of these solitons.

\subsection{Dual of massive ABJM deformation}

To begin, we will compute the spacetime background corresponding to the massive deformation of ABJM. Recall that in the case of pure ABJM, the spacetime background is set up by $N$ $M2$-branes living at the tip of the cone on $R^{2,1}\times C^4/Z_k$. We will deal with the $Z_k$ orbifolding at the end, but for the moment we would like to understand the background that replaces the flat eleven-dimensional $R^{2,1}\times C^4$ in the presence of the massive deformation. To this end, following the suggestion in \cite{Gomis:2008vc} that it could be related to the IIB maximally supersymmetric 
pp wave, we consider the $k=1$ case first and notice that the type IIB gravitational wave 
\bea
  ds^2&=& -dt^2 +dx^2 +(H-1)(dx+dt)^2+d\vec{x}^2\,,\nonumber\\[-0.5cm]
  {}\\
  e^{\phi}&=&g_s\,,\;\;\; B=0\,,\nonumber
\eea
is T-dual\footnote{The easiest way to see this is to use Buscher's T-duality rules \cite{Buscher:1985kb,Buscher:1987sk,Buscher:1987qj} which, in the absence of RR fields are the same for IIA $\rightarrow$ IIB and IIB $\rightarrow$ IIA, namely:
\bea
&&\tilde{g}_{00}=\frac{1}{g_{00}};\;\;\; \tilde{g}_{0i}=\frac{B_{0i}}{g_{00}};\;\;\;
\tilde{g}_{ij}=g_{ij}-\frac{g_{0i}g_{0j}-B_{0i}B_{0j}}{g_{00}}\cr
&&\tilde{B}_{0i}=\frac{g_{0i}}{g_{00}};\;\;\;
\tilde{B}_{ij}=B_{ij}+\frac{g_{0i}B_{0j}-B_{0i}g_{0j}}{g_{00}}\cr
&&\tilde{\phi}=\phi-\frac{1}{2}\ln g_{00}\nonumber
\eea
}
to the fundamental string (F1) solution
\bea
ds^2_{str}&=& H^{-1}(-dt^2+dx^2)+d\vec{x}^2\,,\nonumber\\
B_{tx}&=&1-H^{-1}\,,\\
e^{\phi}&=& H^{-1/2}\,,\nonumber
\eea
where $H=1+Q/r^6$ is a harmonic function of the transverse coordinates $\vec{x}$ and the 
F1 solution is such that all fields are trivial at $\vec{x}\rightarrow\infty$. In turn, using
\bea
ds_M^2&=& e^{4\phi/3}\left(ds_{10}^2+e^{2\phi}
\left(A_\mu dx^\mu+dx_{11}\right)^2\right)\,,\nonumber\\
F_{abcd}^{(M)}&=&e^{4\phi/3}\left(F_{abcd}+4A_{[a}H_{bcd]}\right)\,,\\
F_{abc 11}^{(M)}&=&H_{abc}\,,\nonumber
\eea
the $F1$ solution lifts to the membrane ($M2$) solution 
\bea
  ds_M^2&=&H^{-2/3}d\vec{y}^2+H^{1/3}d\vec{y}^2\,,\nonumber\\
F_4&=& dy^0\wedge dy^1\wedge dy^2 \wedge d H^{-1}\,,\\
H&=&1+\frac{Q}{r^6}\,,\nonumber
\eea
in $M$-theory. Here we define $(y^0,y^1,y^2) = (t,x,y\equiv x_{11})$. On the other hand, the {\it maximally supersymmetric type IIB pp-wave} ,
\bea
ds^2&=& 2dx^+dx^--\mu^2\vec{x}^2\left(dx^+\right)^2+d\vec{x}^2\,,\nonumber\\
F_{+1234}&=&F_{+5678}=\mu\,,\\
e^{\phi}&=&g_s;\;\;\; B=0\,,\nonumber
\eea
can be understood as the gravitational wave obtained in the presence of the constant flux $F_{+1234}=F_{+5678}=\mu$, which also modifies $H-1$ from $Q/r^6$ to $-\mu^2 r^2$. Correspondingly, the 
IIA solution T-dual to this pp-wave is found by using the full IIB $\rightarrow$ IIA Buscher rules (including those for the RR sector):
\bea
\tilde{A}_{ijk}&=&\frac{8}{3}D^+_{ijk}+B_{0[i}B^{(2)}_{jk]}-B^{(2)}_{0[i}B_{jk]}+B_{0[i}B^{(2)}_{|0|j}\frac{g_{k]0}}{
g_{00}}-B^{(2)}_{0[i}B_{|0|j}\frac{g_{k]0}}{g_{00}}\,,\nonumber\\
\tilde{A}_{0ij}&=&\frac{2}{3}B^{(2)}_{ij}+2\frac{B^{(2)}_{0[i}g_{j]0}}{g_{00}}\,,\\
\tilde{A}_i&=&-B_{0i}^{(2)}+aB_{0i};\;\;\; \tilde{A}_0=0\,,\nonumber
\eea
to get
\bea
ds^2_{str}&=& H^{-1}\left(-dt^2+dx^2\right)+d\vec{x}^2\,,\nonumber\\
B_{tx}&=&1-H^{-1}\,,\nonumber\\
e^{\phi}&=& H^{-1/2}\,,\\
F_{1234}&=& F_{5678}=\mu\,,\nonumber\\
H&=&1-\mu^2\vec{x}^2.\nonumber
\eea
This can be interpreted as a fundamental string in the background of the constant flux $F_{1234}= F_{5678}=\mu$, which, again, modifies the harmonic function. Finally, this lifts to the $M$-theory solution
\bea
ds_M^2&=&H^{-2/3}d\vec{y}^2+H^{1/3}d\vec{x}^2\,,\nonumber\\
F_4&=& dy^0\wedge dy^1\wedge dy^2 \wedge d H^{-1}+\mu(dx^1\wedge dx^2\wedge dx^3\wedge dx^4+dx^5\wedge dx^6\wedge
dx^7\wedge dx^8)\,,\\
H&=&1-\mu^2\vec{x}^2\,.\nonumber
\label{massivemsol}
\eea
This can again be interpreted as an $M2$-brane in the presence of constant transverse flux $F_{1234}= F_{5678}=\mu$. This flux (in particular $|F|^{2}$) means that $H$ now satisfies 
$\d_i^2 H=-8\mu^2$ so that, again, the harmonic function is modified. There are two points about this geometry that should be noted: before the T-dualizing, we could always rescale $\mu$ away by sending $x^+\rightarrow x^+/\mu$ and $x^-\rightarrow \mu x^-$. After the T-duality and M-theory lift, this is no longer possible. 
{\it Now} the scale $\mu$ has physical meaning. Secondly, while there is nothing particularly interesting about the point $|\vec{x}|=1/\mu$ in the type IIB metric, here it is potentially singular.

This solution obtained can be interpreted as an $M2$-brane with a constant flux preserving the same manifest $R$-symmetry as the massive deformation of the ABJM field theory. Specifically,
 the $SU(4)_{R}$ is broken to $SU(2)\times SU(2)$, by the splitting of the scalars into the $1,2,3 \& 4$ and $5,6,7 \& 8$ directions. This gives us confidence we are on the right track. For this, and other reasons given in \cite{Gomis:2008vc}, 
we can say with some confidence that we have found the spacetime background corresponding to the massive deformation, at least in the $k=1$ case. For general $k$, we 
still have to apply the $\mathbb Z_k$ action  (inherited from the pure ABJM case) on the transverse coordinates. Since it acts by $Y^A\rightarrow e^{2\pi i/k}Y^A$ the 
solution remains intact and the background is valid at any $k$. 

There is one slight subtlety related to the amount of symmetry preserved under T-duality in the above construction. We started with maximal supersymmetry (32 supercharges, or $\mathcal{N}$=16 in three-dimensions), and ended up with a solution that {\it cannot} have as much since, in M-theory, the {\it unique} backgrounds with maximal susy are flat space, $AdS_4\times S_7, AdS_7\times S_4$ and the maximally supersymmetric eleven-dimensional pp-wave that is their Penrose limit. Since our solution is of the $M2$-brane-type, 
we must have that at least the constraint $\Gamma^0\Gamma^1\Gamma^2\psi=\psi$ hold. This reduces 32 supercharges to 16 (or ${\mathcal N}=8$ in three-dimensions). 
In addition, the presence of the transverse flux $F_{1234}= F_{5678}=\mu$ further reduces the number of supercharges to 12 (giving ${\mathcal N}=6$ in three-dimensions). Naively, this breaks the $R$-symmetry to an $SU(2)\times SU(2)\times U(1)$ which rotates the transverse coordinates $Y^A$. The problem is that in \cite{Gomis:2008vc} it was argued that the massive deformation of the ABJM model retains the full $SO(6)$ $R$-symmetry of N=6 in three dimensions, so the same should apply here. 

On the other hand, in \cite{Gomis:2008cv}, it was argued also that in the BLG (or, equivalently, the $N$=2 ABJM) case, the massive deformation with $M2$ on $R_t \times T^2$ is equivalent to the IIB pp-wave. There, the massive BLG model was observed to satisfy 32 supersymmetries, only 16 of which are linearly realized, and the remaining 16 are nonlinear. The fuzzy sphere background breaks the 16 nonlinear supersymmetries, giving 16 genuine supersymmetries. These fuzzy sphere vacua correspond to $D3$-brane giant gravitons in the IIB pp-wave background, which preserve the same 16 supersymmetries in addition to satisfying the same zero energy condition.

Returning to the case at hand, it seems that the solution to the problem of supersymmetry under T-duality is that one has to consider the full string theory (or rather Matrix theory) and not just the supergravity background. Then one 
needs to consider $D0$-branes in IIB in the presence of the pp-wave. This preserves a total of 16 supercharges, instead of the full 32. After T-dualizing, in the M-theory we have an 
$M2$-brane with constant transverse flux and $H=1-\mu^2\vec{x}^2$, on top of the ABJM 
$M2$-branes. This configuration again preserves 16 of the supercharges, so there is no problem.
Of course, the above argument applies just for the k=1 \& 2 ABJM model (that has no 3/4 supersymmetry reduction), otherwise we need to take a $Z_k$ quotient of the target space.

\subsection{Brane-filling solution: the fuzzy sphere}

The first type of solutions we will analyze in some detail are {\it brane-filling}, i.e. ground states, which can be interpreted as branes with a transverse extension. In \cite{Gomis:2008vc} it was shown that the maximally supersymmetric ground state of the massive deformation of ABJM 
is given by an equation associated with a fuzzy sphere. In particular, the four complex scalars of ABJM as split as $Y^A=(Q^\a, R^{\dot\a})$ and the solution is given by $R^{\dot\a}=0$ and $Q^\a=fG^\a=\sqrt{\frac{k\mu}{4\pi}}$, with
\be
-G^\a=G^\b G^\dagger_\b G^\a-G^\a G^\dagger_\b G^\b\;.\label{GRVV}
\ee 
The irreducible matrices that solve eq.(\ref{GRVV}) - that we will call 
the GRVV algebra - are 
\bea\label{BPSmatrices}
 ( \tilde{G}^1)_{m,n } &=& \sqrt { m- 1 } ~\delta_{m,n}\,,\nonumber\\
 ( \tilde{G}^2)_{m,n} &=& \sqrt { ( N-m ) } ~\delta_{ m+1 , n } \,,\nonumber\\ [-0.5cm]
 {}\\
 (\tilde{G}_1^{\dagger} )_{m,n} &=& \sqrt { m-1} ~\delta_{m,n} \,,\nonumber\\
 ( \tilde{G}_2^{\dagger} )_{m,n} &=& \sqrt { (N-n ) } ~\delta_{ n+1 , m }\,.\nonumber
\eea
Based partly on the fact that $G^\a G^\dagger_\a=N-1$ for this irrep, this solution was conjectured to represent a fuzzy $S^3$, but this was 
shown to not be the case in \cite{Nastase:2009ny}. It is instead a fuzzy $S^2$, as could be guessed by the fact that $G^1=G^\dagger_1$
(so that the scalar represented by the imaginary part of $G^1$ is fixed to zero). Moreover, in \cite{Nastase:2009zu} it was shown that if we mod out by the $U(N)$ gauge 
transformations that leave the GRVV algebra invariant, the resulting bifundamental matrices can be viewed as fuzzy versions of Killing spinors on the 
$S^2$. The action for fluctuations around the solution is then the supersymmetric $D4$-brane action, compactified on $S^2$. The fields can be expanded 
in fuzzy spherical harmonics made up of either $J_i$ or $\bar J_i$,
\bea\label{defining}
J_i&=&{(\tilde\s_i)^\a}_\b G^\beta G^\dagger_\alpha={(\tilde\s_i)^\a}_\b{J^\b}_\a\equiv{(\s_i)_\b}^\a {J^\b}_\a\,,\nonumber\\[-0.3cm]
{}\\
\bar{J}_i&=&{(\tilde\s_i)^\a}_\b G^\dagger_\alpha G^\beta =  {(\tilde{\s}_i)^\a}_\b {\bar{J}_\a\,}^\b\equiv{(\sigma_i)_\b}^\a{\bar{J}_\a\,}^\b\,,\nonumber
\eea
in the same way as the regular spherical harmonics are built from Euclidean coordinates $x_i$. In the classical limit, both the $J_i$ and $\bar J_i$ tend to the same $x_i$, up to a normalization constant. The $D4$-brane compactified on the $S^2$ ground state then corresponds to a $D3$-brane giant graviton wrapping a 3-sphere
\be
\left|Z^1\right|^2+\left|Z^2\right|^2=R^2\;,
\ee
in the maximally supersymmetric type IIB pp-wave. Here the 4+4 coordinates transverse to the IIB pp-wave are $Z^A=(Z^\a,Z^{\dot{\a}})$ with 
\bea
Z^1=X^1+iX^2\;,&\quad& Z^2=X^3+iX^4\,,\nonumber\\[-0.4cm]
{}\\
Z^{\dot 1}=X^5+iX^6\;,&\quad& Z^{\dot 2}=X^7+iX^8\,.\nonumber
\eea
We also have to mod out by $\mathbb Z_k$ which acts on the transverse coordinates as $Z^A\rightarrow e^{2\pi i/k}Z^A$, and $Z^A$ corresponds to $Y^A$ in M-theory.

\subsection{Vortex solutions}
Moving up in co-dimension, in \cite{Arai:2008kv}, a BPS vortex solution was found for another massive deformation of ABJM. This time the deformation is given by 
the superpotential term $\delta W=\mu Tr[{\cal Z}^\a {\cal W}_\a]$, and admits the supersymmetric ground state 
\be
Z^\a=W^{\dagger \a}=\sqrt{\frac{k\mu}{4\pi}}G^\a\,.
\ee
Here we have written $Y^A=(Z^\a,W^{\dagger\dot\a})$ instead of $(Q^\a,R^{\dot\a})$, since the different mass deformation implies a different kind of split. While this deformation manifestly preserves an $SU(2)_R$ symmetry and ${\cal N}=2$ supersymmetry, it was demonstrated in \cite{Kim:2009ny} that it in fact preserves the full ${\cal N}=6$, and it matches the GRVV deformation upon a redefinition of the fields. The BPS vortex solution is given by the ansatz
\bea
 Z^\a&=&W^{\dagger \a}=f(x)G^\a\,,\nonumber\\
 A_\mu&=&a_\mu(x)G^\a G^\dagger_\a\,,\nonumber\\
 \hat A_\mu &=&a_\mu(x)G^\dagger_\a G^\a\,,\\
 f(r,\theta)&=&\sqrt{\frac{k\mu}{4\pi}} g(r)e^{in\theta}\,,\nonumber\\
 a_i(r)&=&\epsilon_{ij}\frac{\hat x_j}{r}[a(r)-n],\;\;\; i=1,2\,,\nonumber
\eea
where $a(r)$ and $g(r)$ are real functions satisfying
\bea
  a(r)&=&\mp rg'(r)/g(r)\,,\nonumber\\[-0.4cm]
  {}\\
  (\ln g^2)''\!\!\!\!&+&\!\!\!\!\frac{1}{r}(\ln g^2)'+4\mu^2 g^2(1-g^2)=0\,,\nonumber
\eea
together with the boundary conditions $a(0)-n=0$, $ng(0)=0$ and $g(\infty)=1$, $a(\infty)=0$.
The vortex has magnetic flux $\Phi=-k\mu n/2$ and zero angular momentum and, at infinity it goes over to the GRVV ground state, i.e. the fuzzy $S^2$. Consequently, at the position of the vortex $r=0$, and $Z^\a=W^{\dagger\a}=0$. It has no extension in the transverse coordinates either (the fuzzy sphere shrinks to zero there), and so can be interpreted as a point in both worldvolume and transverse space directions. This implies it can be identified with an object carrying D0-brane charge\footnote{In \cite{Kim:2009ny} and \cite{Auzzi:2009es}, more general vortex solutions were given, based on a general discussion of the BPS equations of the 
GRVV massive deformation. Because of the identification of the massive deformations up to field redefinitions however, and the fact that vortices
carry topological charge, all the vortex solutions must represent the same object. This soliton has charge $k$ and mass $k\mu$.} \cite{Auzzi:2009es}. The U(1) symmetry 
under which the vortex is charged was identified in the dual gravity background - the $Z_k$ reduction of the LLM solution for $N$ $M2$'s in flat space - as $J=kQ_0+NQ_4$
 where $Q_0$ and $Q4$ are the $D0$-brane and $D4$-brane charges respectively. One interpretation of this solution is as a bound state of an object with 
$D0$-brane charge $k$, the vortex, with the $D4$-brane wrapped on the fuzzy $S^2$ corresponding to the ground state. This picture was confirmed through 
a $D0$-brane probe analysis with the mass of the probe at the minimum matching the soliton mass of $k\mu$. 

Evidently, the vortex truly does represent a charge $k$ $D0$-brane object which, in the large 
$k$ classical limit becomes a classical spacetime object, a supergraviton. It also follows that the vortex solution of the massive ABJM theory is indeed a crucial ingredient in a search for a Matrix model of ABJM.

We note that if the proposal of this paper is correct, we should also find ABJM solutions corresponding to branes, at least $D2$ and $D4$, that are {\it completely transverse}, {\it i.e.} that have a worldvolume in the transverse scalar directions, but are pointlike on the worldvolume. These would be vortex solutions with fuzzy 2- and 4-dimensional spaces respectively, at the position of the vortices. We have, unfortunately, not succeeded in identifying such solutions. 

\subsection{Kink solutions}

According to our proposed Matrix theory-type interpretation for ABJM, we also expect 
to find ABJM solutions that can be interpreted as 2-branes with 
one spatial worldvolume coordinate parallel to the M2-brane and one perpendicular to it. 
Such a solution would be a kink soliton of ABJM model, which at the position of the 
kink can be interpreted as a ``fuzzy circle". The problem however is two-fold: first, we need to find a kink solution of the ABJM model and second, we have to understand what exactly such a ``fuzzy circle" is. Odd-dimensional fuzzy $n$-spheres have proven notoriously subtle to construct in the past when $n\geq 3$ \cite{Ramgoolam:2001zx}. The fuzzy circle, on the other hand, appears not to even have been constructed yet! We will see however that in a particular limit, the fuzzy sphere plays the role of a fuzzy circle.\footnote{A different limit was considered in \cite{Janssen:2004jz},
where a fuzzy circle arose from a "fuzzy cylinder" construction. We thank Yolanda Lozano for pointing this out to us, after the first version of 
our paper was posted on the electronic archives.}

The general problem of finding a fuzzy circle can be readily described. For a fuzzy sphere, the fuzzy spherical harmonics constitute a {\it complete} 
subset of the set of $N\times N$ matrices, with a maximal angular momentum of $N^2-1$.
This also means that the composition of spherical harmonics, $Y_kY_l\sim \sum _n C_{kl}^n Y_n$ is still valid in the 
fuzzy case as long as $k+l\leq L_{max}=N^2-1$. For a fuzzy circle, the ``fuzzy spherical harmonics" (or ``fuzzy Fourier modes") would also need to 
satisfy similar completeness and composition relations. These, however, appear to be quite difficult to obtain in any other way than as a limiting case of the fuzzy sphere. For instance, completeness requires that the $N^2$ elements of the $N\times N$ matrices need to be arranged 
as circle Fourier modes of increasing order $n$ which in turn makes composition hard to accomplish.

We now turn to studying kink solutions of the massive ABJM Lagrangian. Naively, it might seem like a good place to start is with the construction of \cite{Arai:2008kv,Hanaki:2008cu} of the half supersymmetric kink 
\bea
  Z^\a&=&W^{\dagger \a} = f(x_1)G^\a\,,\nonumber\\[-0.4cm]
  {}\\
  f^2(x_1)&=&\frac{k}{4\pi}\frac{\mu}{1+e^{-2\mu x_1}}\,.\nonumber
\eea
However, at the location of the kink, $x_1=0$ and $Z^\a=W^{\dagger\a}=\mu G^\a$, resulting in a degenerate fuzzy  $S^2\times S^2$. This is not the kind of solution we are looking for\footnote{One could perhaps interpret this solution as a 
5-brane with 4 directions transverse to the ABJM worldvolume, but this is less clear}. 

In search of the {\it fuzzy circle kink} solution, we will look for solutions with no gauge fields, and half the scalars set to zero. In the notation of GRVV \cite{Gomis:2008vc}, this corresponds to exciting only the $Q^{\alpha}$ of $Y^{A} = \left(Q^{\alpha},R^{\dot{\alpha}}\right)$. The action on such a solution would be
\bea
  S=\int d^{2}x\,\Biggl[
  \left(\d_\mu Q^\a\right)^2&+&\left[\mu Q^1+\frac{2\pi }{k}\left(Q^1Q_2^  
  \dagger Q^2-Q^2Q^\dagger_2 Q^1\right)\right]^2\nonumber\\
  &+&\left[\mu Q^2+\frac{2\pi }{k}\left(Q^2Q_1^\dagger 
  Q^1-Q^1Q^\dagger_1 Q^2\right)\right]^2\Biggr]\,.
  \label{actionsol}
\eea
We can readily check that the ansatz $\displaystyle Q^1=f(x_{1})G^1\,,\,\, 
Q^2=\sqrt{\frac{-\mu k}{2\pi}}G^2$, where $G^1,G^2$ are matrices satisfying the GRVV fuzzy sphere algebra, results in the following quartic action for the function $f$, 
\be
  S=\frac{N}{2}\int dx_{1}\, \left[\left(\d_x f\right)^2+\frac{-\mu k}{2\pi}\left(\frac{2\pi} 
  {k}f^2 + \mu)\right)^2\right]\,.
\ee
The resulting equations of motion are solved by the usual kink solution $f(x_1)=\tanh x_1$. However, we note that {\it only half of the equations of motion} of the full action eq.(\ref{actionsol}) are solved by this ansatz. This ``near miss" suggests another avenue: Completing the square in the action eq.(\ref{actionsol}) produces the BPS equations
\bea
 \d_s Q^1&=&\mu Q^1+\frac{2\pi }{k}(Q^1Q^\dagger_2Q^2-Q^2Q^\dagger_2Q^1)\,,\nonumber\\[-0.4cm]
 {}\\
 \d_s Q^2&=&\mu Q^2+\frac{2\pi }{k}(Q^2Q^\dagger_1Q^1-Q^1Q^\dagger_1Q^2)\,,\nonumber
\eea
whose solutions will be the half supersymmetric solitons of the theory. The ansatz
$Q^2=fT^2;\;\;\; Q^1=g T^1$, where the $T^i$ are matrices that satisfy the algebra
\bea
  T^1T^\dagger_2T^2-T^2T^\dagger_2T^1&=&-\alpha T^1\,,\nonumber\\[-0.4cm]
  {}\\
  T^2T^\dagger_1T^1-T^1T^\dagger_1T^2&=&-T^2\,,\nonumber
  \label{matrices}
\eea
collapses the BPS equations to two coupled differential equations for $f$ and $g$ whose solution is
\bea
 f(x_1)&=&\sqrt{\frac{-C_1}{1-2\pi \alpha\rm e^{-2C_1C_2/k}\exp(\frac{C_1}{\mu k}\rm e^{2\mu x_1})}}\rm e^{-C_1C_2/k}\exp\left(\mu x_1+\frac{C_1}
 {2\mu k}\rm e^{2\mu x_1}\right)\,,\nonumber\\[-0.4cm]
 {}\\
  g(x_1)&=&\frac{1}{\sqrt{2\pi}}\sqrt{\frac{-C_1}{1-2\pi \alpha\rm e^{-2C_1C_2/k}\exp(\frac{C_1}{\mu k}\rm e^{2\mu x_1})}}\rm e^{\mu x_1}\,.\nonumber
\eea
After fixing one of the integration constants as $C_1=-k^2/2\pi$ we get
\bea 
g&=&\frac{k}{2\pi}\frac{1}{\sqrt{1-\frac{4\pi^2\alpha }{k^2}A^2\exp(-\frac{k}{2\pi\mu}\rm e^{2\mu x_1})}}\rm e^{\mu x_1}\,,\nonumber\\[-0.4cm]
{}\\
f&=&\frac{A}{\sqrt{1-\frac{4\pi^2\alpha }{k^2}A^2\exp(-\frac{k}{2\pi\mu}\rm e^{2\mu x_1})}}\exp\left[\mu x_1-\frac{k}{4\pi\mu}\rm e^{2\mu x_1}\right]\,.\nonumber
\eea
This solution has the following properties: As $x_1\rightarrow\pm \infty$, $f\rightarrow0$. Moreover, $f(x_{1})$ has a maximum at
\be 
x_1=\frac{1}{2\mu}\ln \left[\frac{2\pi \mu}{k}\left(L\left(-\frac{4\pi^2 A^2\alpha}{k\rm e}\right)+1\right)\right]\,,
\ee
of value 
\be 
f_{max}=A\left(\frac{2\pi \mu}{k\rm e}\right)^{1/2}\left[\frac{L\left(-\frac{4\pi^2A^2}{k\rm e}\alpha\right)}{\left(-\frac{4\pi^2A^2}{k\rm e}\alpha\right)}\right]^{\frac{1}{2}}\,,
\ee
where $L(\xi)$ is a Lambert W function that satisfies the Lambert equation,
\begin{eqnarray}
  z\left(1+L\right)\frac{dL}{dz} = L\,,\quad z\neq -\frac{1}{e}\,.
\end{eqnarray}
This solution can be interpreted as a (non-topological) kink with regard to the field 
$f$, while nothing particularly interesting happens for the field $g$. As for the transverse fuzzy circle, note that the $T^\a$ can be taken to be 
\be
T^1=G^1;\;\;\; T^2=\sqrt{\a}G^2\,,
\ee
and hence define deformed fuzzy spheres. Up to now, the introduction of the parameter $\a$ seems to be superfluous. However, note that if we take the 
$\a\rightarrow 0$, then what we have is a BPS solution with $T^1=G^1$ and $T^2=0$, yet satisfying the nontrivial relations\footnote{Of course, if $T^2=0$ from the start, these equations are trivial. They are only nontrivial on taking the limit of $\a\to 0$.}
\bea
&&T^1T^\dagger_2T^2-T^2T^\dagger_2T^1=0\,,\nonumber\\[-0.4cm]
&{}&\\
&& T^2T^\dagger_1T^1-T^1T^\dagger_1T^2=-T^2\,.\nonumber
\eea
This is ample justification for our associating them with the fuzzy spherical harmonics of the fuzzy circle. In this limit, the solution becomes
\bea 
g&=&\frac{k}{2\pi} e^{\mu x_1}\,,\nonumber\\
&{}&\\
f&=&A\exp\left[\mu x_1-\frac{k}{4\pi\mu}\rm e^{2\mu x_1}\right]\,,\nonumber
\eea
and is still a kink, since $f$ has the same asymptotics with a maximum now at 
\be
x_1=\frac{1}{2\mu}\ln\left(\frac{2\pi\mu}{k}\right),
\ee
while the field $g$ exhibits a simple monotonic increase. The interpretation of this result is that we have taken a fuzzy sphere and shrunk one of its directions ($G^2$) to zero,
thus obtaining a fuzzy circle\footnote{As a final point, note that our choice of representation for the $G$'s is not unique. We could take for $G^1$ and $G^2$ either the irreducible GRVV matrices (as we have), or the reducible matrices $G^1=\delta_{m+N/2,n}$, $G^\dagger_1=\delta_{m, n+N/2}$ and $G^2=\delta_{mn}$ with $m,n\leq N/2$.}.



\section{Graviton scattering in massive ABJM background}

There are two cases of interest for the scattering of gravitons in the background given in eqs.(5.8). These correspond to the two types of vortex scattering in the massive ABJM model that we want to match to. There, the two relevant cases are when the vortices are separated in 
the directions parallel to the ABJM worldvolume, or in the transverse (scalar) directions. 

In one case the gravitons are separated only in the two directions parallel to the ABJM worldvolume, and the 11th dimension is transverse to the 
worldvolume. In the second, they are separated only in the directions transverse to the ABJM worldvolume, and the 11th dimension is parallel to it. This latter case is a more direct analog of the BFSS analysis which we reviewed previously, whereas the former case is new. 
As in the BFSS model, we will employ 't Hooft's trick of replacing the graviton-graviton scattering with the scattering of a graviton shockwave with a probe wave, but here, it is not clear {\it a priori} that it will give the complete result.

As in the BFSS case, the graviton shockwaves satisfy the wavefunction equation, 
\be
\Delta^\perp_{bgr}h_{--}=Q\delta^\perp(\vec{x},\vec{y})\,.
\ee
We will content ourselves with some cursory comments about the solutions of this equation here and leave a detailed analysis of the solutions for the Appendix. As earlier, we will denote the worldvolume coordinates by $\vec{y}$ (with $|\vec{y}|\equiv y$), and transverse coordinates by 
$\vec{x}$ (with, correspondingly, $|\vec{x}|\equiv r$). In the transverse separation case, we find 
\be
h_{--}(\vec{y}=0,r)\sim C' \frac{Q}{r^7}\,,
\ee
as in the BFSS case, but since the result is independent of $\mu$, we could have imposed the wrong boundary condition, so we will refrain from using this result in the following. In the case of parallel separation on the other hand, we find that
\be
h_{--}(\vec{y},r=0)\sim C e^{-m_1 |y|}\,.
\ee
The next step in this computation is to scatter a scalar probe-graviton off the source-graviton. Since it is clearer how to deal with the transverse case, we will do this first, more as a guide than anything else. 

\noindent
{\bf 1. Transverse separation:} Expanding the D'Alembertian in the probe graviton equation 
\be
\left[\Box^{full}_{bgr}+\d_+ h^{++}\d_+\right]\phi=0\,,
\ee
gives
\be
H^{-1/3}\left[H(\vec{x})\left(\d_y^2-\frac{\d^2}{\d x_0^2}+m^2\right)+\d^2_{\vec{x}}\right]\phi=\left[-\frac{h^{++}}{2}(\d_0-m)^2\right]\phi
\equiv V\phi\,.
\ee
As in the, perhaps more familiar, BFSS case the momentum-space interaction potential is computed through the S-matrix
\be 
S^{(1)}(p_{in}-p_{out})=\int d^{10}z\,\,\, \phi_{out}^{(0)}(z)^*V(z)\phi_{in} ^{(0)}(z)\,,
\ee
where $\phi^{(0)}$ solves the unperturbed equation
\be
\left[H(\vec{x})\left(\d_y^2-\frac{\d^2}{\d x_0^2}+m^2\right)+\d^2_{\vec{x}}\right]\phi^{(0)}=0\,.
\ee
If we factorize the plane wave $\phi^{(0)}$ as $\phi^{(0)}=\frac{e^{ip_yy-iEx_0}}{\sqrt{2E}}f(\vec{x})$, define $p_0=\sqrt{E^2+m^2-p_y^2}$, $\vec{y}=\sqrt{\mu p_0}\vec{x}$ and 
$\tilde{E}=(p_0)^2/(2\mu p_0)$, we find that the function $f$ satisfies
\be
 \left[\d_{\vec{y}}^2-
 \vec{y}^2+2\tilde{E}\right]f=0\,,
 \label{feq}
\ee
which is nothing but the equation of a $d$-dimensional harmonic oscillator. Therefore, if we were interested in solutions that drop off at infinity,
the solution would be given in terms of Hermite polynomials as
\bea
  f&=&\prod_i f_{p_0,i};\;\;\;\;
  f_{p_0,i}=\frac{e^{-y_i^2/2}H_{n_i}(y_i)}{\pi^{1/4}\sqrt{2^{n_i}n_i!}}\,,
  \nonumber\\[-0.4cm]
  &{}&\\
  \tilde{E}&=&\sum_i\tilde{E}_i\,,\nonumber
\eea
with the quantization condition $\tilde{E}_i=n_i+1/2$.

However, we have $|\vec{x}|\leq 1/\mu$ and, as explained in the Appendix, we need to impose Neumann boundary conditions there instead. Consequently, the ``plane wave solution" of (\ref{feq}) is 
\bea
  f&=&\prod_i f_{p_0,i}\,,\nonumber\\
  f_{p_0,i}&=&e^{-y_i^2/2}\left[C_{1,i}\,\,\,{}_{1}F_{1}\left(\frac{1}{4}-
  \frac{\tilde{E}_i}{2},\frac{1}{2};y_i^2\right) +C_{2,i}\,\,y_i\,\,
  \,{}_{1}F_{1}\left(\frac{3}{4}-\frac{\tilde{E}_i}{2},\frac{3}{2};y_i^2\right)\right]\,,\nonumber\\
  &{}&\\
  \frac{(p_0)^2}{2\mu p_0}&=&\tilde{E}=\sum_i\tilde{E}_i\equiv 
  \frac{(p_{0,i})^2}{2\mu p_0}\,,\nonumber\\
  y_i&=&\sqrt{\mu p_0} x_i\,.\nonumber
\eea
Here the relative normalization $C_1/C_2$ {\it could} be defined by the 
Neumann boundary condition at $r=1/\mu$, but that would result in {\it two constants} for each dimension. We will see in what follows that another relative normalization is preferred. 
Note that because $p_0^2=\sum_i (p_{0,i})^2$, $\vec{p}_0$ can be called the (plane wave) momentum, and we will fix the normalization by requiring that we obtain the flat space plane waves in some limit. 

Indeed, from Appendix C of \cite{Jevicki:2005ms}, we know that the harmonic oscillator wavefunctions become plane waves as, for example $n\rightarrow\infty$ or 
$m\omega\rightarrow 0$. Explicitly,
\bea
\phi_n(x)&=&\frac{(m\omega)^{1/4}}{\left(\sqrt{\pi}2^n n!\right)^{1/2}}e^{-m\omega x^2/2}H_n\left(\sqrt{m\omega}x\right)
\rightarrow  \frac{1}{\sqrt{\pi}}\left(\frac{4 m \omega}{n}\right)^{1/4}\cos\left(2\sqrt{m\omega n}x\right)
\eea
where $k_n=\sqrt{2 m \omega n}=2\pi n/L$. This limit, together with the relations
\bea
H_{2n}(x)&=&(-1)^n\frac{(2n)!}{n!}\,{}_{1}F_{1}\left(-n, \frac{1}{2};x^2\right)\,,\nonumber\\[-0.4cm]
&{}&\\
H_{2n+1}(x)&=&(-1)^n\,2\,\frac{(2n+1)!}{n!}\,x\,\,{}_{1}F_{1}\left(-n,\frac{3}{2};x^2\right)\,,\nonumber
\eea
generalize to the limit of our solutions, giving plane waves for large $p_{0,i}/\mu$. We fix the relative constants
$C_{1i}/C_{2i}$ such that we are able to sum $\cos(\cdots)+i\sin(\cdots)=e^{i(\cdots)}$ after the limit. As we will argue further on, knowledge of the precise values of these constants is not necessary, only that the plane wave limit exists. The overall constant in front of the solution follows from a relativistic normalization.

We now reverse the initial logic, take the $p_{0,i}$ to be independent, and instead constrain $E$ by the relation $E^2+m^2-p_y^2=p_0^2=\sum_i (p_{0,i})^2$. The interaction potential is then obtained from
\bea 
S^{(1)}(E;p_{in}-p_{out})&=&\!\!\int dx_0\; dy\; d^8\vec{x}\; \overline{\phi_{out:p_y,m,\vec{p}_{0}}^{(0)}} \left[-\frac{h^{++}(y,\vec{x})}{2}(\d_0-m)^2\right]
\phi_{in:p_y,m,\vec{p}_0} ^{(0)}\nonumber\\
&=&-\frac{2\pi\delta(E^{out}-E^{in})(E^{in}-m)^2}{4\sqrt{E^{in}E^{out}}}
\int\!\! dy\;e^{iy(p_y^{in}-p_y^{out})}\int\!\! d^8\vec{x}\,\,
\overline{f^{out}_{\vec{p}_0^{out}}(r)}h^{++}(y,r)f^{in}_{\vec{p}_0^{in}}(r)\nonumber\\
&=&2\pi \delta(E^{out}-E^{in})V_{int}^{(1)}(E;p_{in}-p_{out})\,.
\eea
with the understanding that we need to make a Fourier transform back to $x$ space at the end. 
However note that since $f_{\vec{p}_0}(\vec{x})\neq e^{i\vec{p}_0\vec{x}}$, it is not even guaranteed that we get a function of $|\delta\vec{p}_0|=|\vec{p}_0^{in}-\vec{p}_0^{out}|$ 
after the $\vec{x}$ integration. Assuming that it is nevertheless
such a function, after the Fourier transform back to position space we find that
\bea
V_{int}^{(1)}(y,\vec{z})&=&\frac{(E-m)^2}{2E}\int d^8\delta\vec{p}_0\,\, e^{-i\vec{z}\delta\vec{p}_0}\int d^8 \vec{x}\,\,
\overline{f^{out}_{\vec{p}_0^{out}}(\vec{x})}h^{++}(y,r)f^{in}_{\vec{p}_0^{in}}(\vec{x})\cr
&\simeq& \frac{mv^4}{8}\int d^8\delta\vec{p}_0\,\, e^{-i\vec{z}\delta\vec{p}_0}\int d^8 \vec{x}\,\,
\overline{f^{out}_{\vec{p}_0^{out}}(\vec{x})}h^{++}(y,r)f^{in}_{\vec{p}_0^{in}}(\vec{x})
\eea
In particular, when $y=0$,
\be
V_{int}^{(1)}(y=0,\vec{z})=\frac{mv^4}{8}\int d^8\delta\vec{p}_0\,\, e^{-i\vec{z}\delta\vec{p}_0}\int d^8 \vec{x}\,\,
\overline{f^{out}_{\vec{p}_0^{out}}(\vec{x})}h^{++}(y=0,r)f^{in}_{\vec{p}_0^{in}}(\vec{x})
\label{transvresult}
\ee
While it is not clear whether the integral is dominated by the low $r$ region,
what is clear is that we cannot have a result depending on the $\delta{p}_0$ unless we are in the plane wave limit. This is obtained in the 
limit of high momentum $p_{0i}$. {\it It seems then that  only if we probe with sufficiently high momentum we will have a well-defined 
interaction potential}.

In any case, the outcome is that we need to replace the functions $f_{\vec{p}_0}$ with plane waves and obtain, as in the BFSS flat space case,
\be
V_{int}^{(1)}(y=0,\vec{z})=\frac{mv^4}{8}h^{++}(y=0,r)\,.
\ee

\noindent
{\bf 2. Parallel separation:}\\
To write down the equation for a probe graviton, we need to first define the coordinates and metric more a little more carefully. We start with the metric
\bea
  ds^2&=&H^{-2/3}(\vec{x})(-dt^2+d\vec{y}^2)+H^{1/3}d\vec{x}^2\,,\nonumber\\[-0.4cm]
  &{}&\\
  d\vec{x}^2&=&dr^2+r^2d\Omega_7^2;\;\;\;\; \vec{x}=(x^1,...,x^8)\,,\nonumber
\eea
and dimensionally reduce on $x_{11}$. Operationally, we choose $x_{11}$ as the fibre direction of 
the Hopf fibration of $S^7$ over $CP^3$ which acts on the Euclidean coordinates, $Y^{A}$, on the $S^{7}$ through the overall phase $Y^A = e^{ix_{11}}\tilde Y^A$. Since the coordinates satisfy $|Y^A|^2/r^2=1$, this means that if we write $\vec{x}=(\vec{x}',x_{11})$, then $r=|\vec{x}|=|\vec{x}'|$ and $d\vec{x}^2=|dY^A|^2=|d\tilde Y^A|^2+dx_{11}^2$ becomes $d\vec{x}'^2+dx_{11}^2$. Since the metric is translationally invariant in $x_{11}$, we can schematically rewrite 
\be
ds^2\simeq H^{-2/3}(\vec{x}')\left(-dt^2+H(\vec{x}')dx_{11}^2+d\vec{y}^2\right)
+H^{1/3}(\vec{x}')d\vec{x}'^2\,.
\ee
With this metric, the equation for the probe graviton takes a form similar to the transverse separation case,
\be
H^{-1/3}\left[H(\vec{x})\left(\d_{\vec{y}}^2-
\frac{\d^2}{\d x_0^2}\right)+\Delta\left(\vec{x}',m^2\right)\right]\phi=\left[-\frac{h^{++}}{2}(\d_0-m)^2\right]\phi\,,
\equiv V\phi
\label{parinteq}
\ee
where $\Delta\left(\vec{x}',m^2\right)$ is obtained from $\d_{\vec{x}}^2$ by reducing on $x_{11}$, and setting
$\phi(x_{11},...)=e^{ipx_{11}}\phi(...)$ and $p=im$. Noting that we could rewrite $-dt^2+H(\vec{x}')dx_{11}=2dx^+dx^-$ only on slices of (nearly) constant $\vec{x}'$, we picked the simplest choice $\vec{x}'\simeq 0$ and fixed it so that we can form the $x^+$ combination for 
defining $h^{++}$ without problem. 

Now we can proceed as with the transverse case. Specifically, the momentum-space 
interaction potential obtains from
\be 
S^{(1)}(p_{in}-p_{out})=\int d^{10}z\,\, \phi_{out}^{(0)}(z)^*V(z)\phi_{in} ^{(0)}(z)\,,
\ee
where $\phi^{(0)}$ satisfies the unperturbed equation associated to eq.(\ref{parinteq}). Here
again, a plane wave ansatz of the form $\phi^{(0)}=\frac{e^{i\vec{p}_y\vec{y}-iEx_0}}{\sqrt{2E}}f(\vec{x})$ will solve the equation provided
\bea
\left[H(\vec{x})\left(E^2-\vec{p}_y^2\right)+\Delta\left(\vec{x}',m^2\right)\right]f(\vec{x})=0\,.
\label{intermedia}
\eea
We again obtain
\bea 
S^{(1)}(E;p_{in}-p_{out})&=&\!\!\!\int dx_0\; d^2y\; d^7\vec{x}\; \overline{\phi_{out:\vec{p}_y,m,\vec{p}_0}^{(0)}}\left[-\frac{h^{++}(\vec{y},\vec{x})}{2}(\d_0-m)^2\right]
\phi_{in:\vec{p}_y,m,\vec{p}_0}^{(0)}\nonumber\\
&{}&\\
&=&\!\!-\frac{\delta(E^{out}-E^{in})(E^{in}-m)^2}{4\sqrt{E^{in}E^{out}}}
\int d^2y\;e^{i\vec{y}(\vec{p}_y^{in}-\vec{p}_y^{out})}\int d^7\vec{x}\,\,
\overline{f^{out}_{\vec{p}_0^{out}}}
h^{++}(y,r)f^{in}_{\vec{p}_0^{in}}\nonumber
\eea
or, after manipulations similar to the transverse separation case,
\be
V_{int}^{(1)}(\vec{y},\vec{z}=0)\simeq\frac{(E-m)^2}{4E}\Big[\int d^7 \delta\vec{p}_0\int d^7\vec{x}
f^{out}_{\vec{p}_0^{out}}(\vec{x})^*h^{++}(y,r)f^{in}_{\vec{p}_0^{in}}(\vec{x})\Big]\,.
\label{parallelresult}
\ee
We have, of course, already noted that equations (\ref{parinteq}) through (\ref{intermedia}) 
only make sense if $\vec{x}\simeq 0$ is implied, in order to be able to form the $x^+$ combination. This could be 
alleviated by integrating $\vec{x}$ only in a neighbourhood of zero, in which case
\be
V_{int}^{(1)}(\vec{y},\vec{z}=0)\simeq\frac{(E-m)^2}{4E}h^{++}(y,r=0)\left[\int d^7 \delta\vec{p}_0\int_{|x|\simeq 0} 
d^7\vec{x}\,\,\overline{f^{out}_{\vec{p}_0^{out}}(\vec{x})}f^{in}_{\vec{p}_0^{in}}(\vec{x})\right]\,.
\ee
In both cases however, we still have the same problem that we observed in the transverse separation case. Namely, in order for the momentum 
space result to only be a function of $\delta \vec{p}_0=\vec{p}_0^{in}-\vec{p}_0^{out}$, we need to only consider high momentum wavefunctions, which are effectively just plane waves. In that case, the integrations disappear, irrespective of whether or not $\vec{x}\simeq 0$, and we obtain 
\be
V_{int}^{(1)}(\vec{y},\vec{z}=0)\simeq C\frac{(E-m)^2}{4E}h^{++}(y,r=0)\,,
\ee
or, since $h^{++}\propto e^{-m_1y}$,  
\be
V_{int}^{(1)}(\vec{y},\vec{z}=0)\propto m v^4 e^{-m_1 y}\,.
\ee
So much for the graviton-graviton scattering.

\section{Vortex scattering in massive ABJM}

In this section\footnote{This section was done in collaboration with Toshiaki Fujimori. We thank him for allowing its use in this paper.} we analyze vortex scattering in the $N=2$ (or equivalently, the $U(2)\times U(2)$) massive ABJM model. The general $N$ case is quite 
complicated, so we will not attempt it here. We will employ the vortex solution in the form of \cite{Kim:2009ny}, as it is both guaranteed to be 
the most general solution\footnote{since it was not found using any specific ansatz, but rather by analyzing the energy functional on a case-by-case basis.},
and also because the specific form of the solution will allow the use of previous known results for the Abelian-Higgs model and its Nielsen-Olesen vortex solution. The vacuum for the massive $U(2)\times U(2)$ ABJM model is 
\be
Y^1=\sqrt{\frac{k\mu}{\pi}}G^1;\;\;\;\;
Y^2=\sqrt{\frac{k\mu}{\pi}}G^2;\;\;\;
Y^3=Y^4=0\,,
\ee
so that, keeping $Y^3=Y^4=0$ and using complex notation for $z=x^1+ix^2$ means that in terms of 
\bea
&&{\cal D}_z Y^\a=\partial Y^\a+iA_zY^\a-iY^\a\hat A_z;\;\;\;
{\cal D}_{\bar z} Y^\a=\bar\partial Y^\a+iA_{\bar z}Y^\a-iY^\a\hat A_{\bar z}\,,
\nonumber\\[-0.4cm]{}\\
&&A_z=\frac{1}{2}(A_1-iA_2);\;\;
A_{\bar z}=\frac{1}{2}(A_1+iA_2);\;\;\;
\hat A_z=\frac{1}{2}(\hat A_1-i\hat A_2);\;\;
\hat A_{\bar z}=\frac{1}{2}(\hat A_1+i\hat A_2)\,,\nonumber
\eea
the BPS equations read
\bea
&&{\cal D}_{\bar z}Y^1=0;\;\;\;
{\cal D}_1 Y^2={\cal D}_2Y^2=0\,,\nonumber\\
&&{\cal D}_0 Y^1-i(\beta^{21}_2+\mu Y^1)=0;\;\;\;
{\cal D}_0 Y^2+i(\beta^{12}_1+\mu Y^2)=0\,,\\
&&\frac{k}{2\pi}\epsilon^{\mu\nu\rho}F_{\nu \rho}=i(Y^\a(D^\mu Y^\a)^\dagger-D^\mu Y^\a Y^\dagger_\a);\;\;\;
\frac{k}{2\pi}\epsilon^{\mu\nu\rho}\hat F_{\nu\rho}=i((D^\mu Y^\a)^\dagger Y^\a-Y^\dagger_\a D^\mu Y^\a)\,\nonumber
\eea
where $\a=1,2$ and
\be
\beta^{\a\b}_\g=\frac{4\pi}{k}Y^{[\a}Y^\dagger_\g Y^{\b]}\,.
\ee
The multi-vortex solution can then be written as 
\bea
&&Y^1=e^{-\frac{\psi}{2}}H_0(z)\sqrt{\frac{k\mu}{2\pi}}\begin{pmatrix} 0&1\\ 0&0\end{pmatrix};\;\;\;
Y^2=\sqrt{\frac{k\mu}{2\pi}}\begin{pmatrix}0&0\\  0&1\end{pmatrix}\nonumber\\[-0.4cm]
{}\\
&&A_0=\frac{1}{\mu}\begin{pmatrix}\d\bar\d \psi &0\\ 0 &0\end{pmatrix};\;\;\;
\hat A_0=\frac{1}{\mu}\begin{pmatrix}0&0\\ 0&\d\bar\d \psi\end{pmatrix};\;\;\;
A_{\bar z}=\hat A_{\bar z}=\begin{pmatrix}0&0\\0&\frac{i}{2}\bar \d\psi\end{pmatrix}\,.\nonumber
\eea
Here $H_0(z)$ is an arbitrary polynomial
\be
H_0(z)=\prod_{i=1}^n(z-z_i)\,,
\ee
and the function real $\psi(z, \bar z)$ is determined through the equation
\be
\d\bar\d \psi=\mu^2\left(1-e^{-\psi}\left|H_0(z)\right|^2\right)\label{psiH}\,,
\ee
with the boundary condition at $|z|\rightarrow \infty$ requiring $\psi\rightarrow \log |H_0(z)|^2$, and where $z_i$ are the position moduli for $n$ vortices.
Correspondingly, the energy of this solution is $nk\mu$. 
Note that this is the same equation governing the vortices in the Abelian-Higgs model, so we expect that the same effective action governs their
scattering as well. This is indeed the case. 

To obtain the effective action for vortex scattering, we let the parameters $z_i$ become functions of $t$, so that the fields of the static 
solution become a function of $(z,\bar z, z_i(t), \bar z_i(t))$. Of course, if we do that, the equations of motion are no longer satisfied, but 
we can still solve them order-by-order in time derivatives $\d_t\sim \dot z_i(t)$. We write
\be
Y^\a(z,\bar z, t)=Y^\a_{(0)}\left(z,\bar z,z_i(t),\bar z_i(t)\right)+Y^\a_{(1)}+Y^\a_{(2)}+...
\ee
and similarly for $A_\mu, \hat A_\mu$. Because they are multiplied by the zero-th order equations of motion, the second order fields do
not contribute to the effective action, so we can stop at the first order solution. At that level  (on-shell),
\be
L_{eff}=\int d^2 x{\cal L}(Y^\a,A_\mu, \hat A_\mu)\,.
\ee
The first order solution is found to be 
\bea
&&Y^\a_{(1)}=0;\;\;\; A_0^{(1)}=\hat A_0^{(1)}=\begin{pmatrix} 0&0\\0&-\frac{i}{2}(\dot z^i\d_i-\dot{\bar z}^i\bar \d_i)\psi\end{pmatrix}\nonumber\\\,,
{}\\
&&A_{\bar z}^{(1)}=\begin{pmatrix} \frac{1}{2\mu}\dot z^i\d_i\d \psi &0\\0&0\end{pmatrix};\;\;\;
\hat A_{\bar z}^{(1)}=\begin{pmatrix}0&0\\0&\frac{1}{2\mu}\dot z^i\d_i\d \psi\,.
\end{pmatrix}
\nonumber
\eea
Finally, substituting this solution into the Lagrangian, produces an effective Lagrangian for the moduli
\be
L_{eff}=\frac{k\mu}{\pi}\int d^2x \left[-\d\bar\d\psi+\frac{1}{2}\dot z^i\dot{\bar{z}}^j
\left(\d_i\bar\d_j \psi+\frac{1}{\mu^2}(\d_i\bar\d\psi\bar\d_j\d\psi-\d\bar\d\psi\d_i\bar\d_j\psi)\right)\right]\,,
\ee
where we have used the equation of motion (\ref{psiH}) and its boundary condition at infinity. This effective Lagrangian is exactly the same as the one obtained
in the Abelian Higgs model for the Nielsen-Olesen vortex! So both the equation of motion (\ref{psiH}) and the effective action for the nonabelian 
Chern-Simons ABJM model give the same result as the ones for the Abelian Higgs model. We can then rewrite the effective Lagrangian for large vortex separation as in the Abelian Higgs model as 
\be
L_{eff}\simeq -k \mu n+\sum_{i=1}^n\frac{k\mu}{2}|\dot{z}^i|^2-k\mu q \sum_{i>j}K_0(2\mu|z^i-z^j|)|\dot{z}^i-\dot{z}^j|^2, \;\; (q\simeq 1.71)
\ee
which means that at large separation, the interaction potential becomes
\be
V_{int}\simeq -k\mu q K_0(2\mu y)v^2\simeq -k\sqrt{\mu} q \frac{\sqrt{\pi}}{2}\frac{e^{-2\mu y}}{\sqrt{y}}v^2
\ee

\section{Towards a Matrix model of massive ABJM?}

Since we have now computed the scattering of both supergravitons and vortices in ABJM, we can compare them to see if they match. Assuming that they play the role of supergravitons\footnote{The background of D4 on fuzzy $S^2$ is common to both vortices in the scattering problem and thus does not contribute to the interaction}, the effective potential for vortices was found to be  $\propto v^2 e^{-2\mu y}$, with $y$ the parallel separation. The $v^2$ dependence
is a general characteristic of lowest order soliton scattering, and the $e^{-2\mu y}=e^{-m_0 y}$ arises from the presence of a lowest-mass excitation 
in the ``broken" phase, i.e. the fuzzy sphere background at infinity. By contrast, we have found that the supergraviton result is proportional to $v^4e^{-m_1 y}$. Here, the $v^4$ was due to the same calculation as in the BFSS case while $m_1\simeq 9.1\mu$ appeared 
because of the necessary Neumann condition at $r=1/\mu$ which, in effect compactifies the transverse direction, with $m_1$ being the lowest KK-mode. Unfortunately then, the interaction potentials for separation in the directions parallel to the ABJM worldvolume do not seem to match.

One possible reason for this mismatch - assuming, of course, that our identification of the vortices as supergravitons is at least correct - appears when we realize that the same $v^4$ dependence appeared on the gravity side as in the BFSS model. Unlike in BFSS where the first interaction term is a one-loop effect, however, here on the matrix side we computed a {\it classical effect}. Clearly then, we need to think a little harder about what terms are supposed to match on both sides, and try to find a corresponding gravity term for the vortex calculation we did, and a corresponding vortex term for the gravity calculation.

There is potentially one other subtlety, though {\it a priori}, it seems to give only small corrections, namely that the gravity background we took is not 
complete. The pure ABJM field theory corresponds to the IR of $M2$-branes on $C^4/Z_k$, but we also need to consider an appropriate vacuum. For pure ABJM, the 
vacua are just VEV's which correspond to a given position $z^I$ in $C^4/Z_k$. On the other hand, in the massive ABJM theory the same VEVs, corresponding to positions $z^I$ in the dual gravity background eqs.(5.8), are no longer solutions. Instead, the vacua are fuzzy spheres of radius $r=1/\mu$, giving an $M5$-brane wrapped on $S^3/Z_k$, which will backreact and 
modify the background, maybe smoothing out the $r=1/\mu$ potential singularity. 

Before we continue trying to fix the mismatch, it will be useful to build some intuition for the problem by looking in more detail at the matching in the BFSS model.

\subsection{BFSS matching: details}

The review of the systematics of Matrix theory matching in this section will follow closely \cite{Becker:1997xw}. With the rescalings $\tau=u/R$ and $X^i=y^i/M_P^3$, the 
bosonic part of the BFSS Matrix theory action (\ref{BFSSaction}) becomes 
\be
S=\frac{1}{M_P^6}\int du\, \Tr\Big(\frac{1}{2}D_uy^iD_uy^i+\frac{1}{4}\left[y^i,y^j\right]^2\Big)\,,
\ee
making explicit the fact that the coupling of the theory is $M_P^3$ (or that the loop-counting parameter is $M_P^6$). At $L$ loops then, the BFSS effective action takes the schematic form 
\be
S_L=M_P^{6L-6}\int du\;f_L\left(y^i,D_u\right)=RM_P^{6L-6}\int d\tau f_L\left(M_P^3X^i,R^{-1}D_\tau\right)\,,
\ee
and, as advertised in section 3, there are two sources of $M_P$-dependence in the interaction potential; one coming from the coupling constant and the other from the use of $M_P$ to build a unit of length. Further analysis, including keeping only Lorentz invariant terms, gives the result
\be
S_L=\int \!d\tau\,\, RM_P^2\left(M_Pr\right)^{4-3L}\left(\frac{v^2}{R^2M_P^2(M_Pr)^4}\right)^n
\ee
so that, for the first few terms
\bea
{\cal L}_0=&\frac{c_{00}}{R}v^2\nonumber\\
{\cal L}_1=&0&+\frac{c_{11}}{M_P^9R^3}\frac{v^4}{r^7}+\frac{c_{12}}{M_P^{15}R^5}\frac{v^6}{r^{11}}+...\nonumber\\[-0.4cm]
&{}&\\
{\cal L}_2=&0&+\frac{c_{21}}{M_P^{12}R^3}\frac{v^4}{r^{10}}+\frac{c_{22}}{M_P^{18}R^5}\frac{v^6}{r^{14}}+...\nonumber\\
{\cal L}_3=&0&+\frac{c_{31}}{M_P^{15}R^3}\frac{v^4}{r^{13}}+\frac{c_{32}}{M_P^{21}R^5}\frac{v^6}{r^{17}}+...\nonumber
\eea
Notice that the diagonal terms in this series that contribute to the classical general relativity scattering themselves receive contributions {\it from all Matrix theory loops}. 
The Einstein action has an infinite number of vertices, thus tree level scattering gives an infinite series, but they all come with integer powers of $M_P^9\sim 1/G_{11}\sim 1/(\kappa_{11})^2$. On the other hand, the off-diagonal terms 
in the Matrix theory loop expansion must come from quantum correction to the effective action of gravity, since they have 
noninteger powers of $\kappa_{11}^2$. 

As an example, a source-probe approximation for the true classical general relativity interaction potential of gravitons gives
\be
{\cal L}=\frac{N_2}{2R}v^2+\frac{15}{16}\frac{N_1N_2}{R^3M_P^9}\frac{v^4}{r^7}+\frac{225}{64}\frac{N_1^2N_2}{R^5M_P^{18}}
\frac{v^6}{r^{14}}+{\cal O}\left(\frac{v^8}{r^{21}}\right)\,,
\ee
where the result is linear in the probe momentum $N_2/R$ because of the approximation, and $N_1/R$ is the source momentum. The symmetrization of the last term matches against a 2-loop Matrix theory result, 
\be
\frac{225}{32}\frac{1}{R^5M_P^{18}}\frac{v^6}{r^{14}}\frac{N_1N_2^2+N_1^2N_2}{2}\,.
\ee
In any case, we see that in BFSS, all the dependence on $v,r,M_P$ and $R$ could be determined from general principles. It seems pertinent to ask if the same could not be done for ABJM? The answer it seems is {\it no}, at least not for the pure ABJM theory which is conformal. There, all $M_P$-dependence is tied to the $r$-dependence so we cannot take advantage 
of having a dimensionful rescaled field and coupling, like $y^i$ and $M_P^3$, to use in dimensional analysis. Perhaps in the massive ABJM that we used for calculations something can be done since, for instance, the size of the fuzzy sphere vacuum $f=\sqrt{k\mu/(4\pi)}$ sets a dimensionful scale. This remains an open issue. Sadly then, it appears that the intuition of BFSS matching is not particularly applicable to our case.

\subsection{ABJM matching and effective Lagrangian}

Let's step back for a minute and take stock. Our analysis up to this point has not succeeded in producing a concrete Matrix theory for the ABJM model. However, we have learnt a significant amount about the structure of the problem that we are in a position to speculate on what 
the sought after Matrix theory and its matching to gravity would look like. To outline the way forward as clearly as possible, we will focus our speculations on three specific points; 
the matching of parameters on both sides, what calculations can be done on the ABJM gauge theory, and what on the gravity side.

\begin{itemize}
\item {\bf Parameters:} The parameters available on the gravity side of the calculation
(for the background dual to the massive deformation of ABJM) are: the mass deformation $\mu$, the inverse size of the 11th direction $1/R$, the Planck scale $M_P$, 
the source and probe 11th momenta $m_s=N_s/R$ and $m_p=N_p/R$ and $m_1$, the effective mass of the graviton wavefunction $h_{--}$ in the presence of the Neumann condition at $r=1/\mu$. In the souce-probe calculation, the source graviton had $Q\propto m_s$ and the probe graviton had mass $m=m_p$. 

Moving to the massive ABJM field theory, $\mu$ is the same deformation parameter. The integer 
$k$ can be used to define the $1/R$ scale with respect to some other scale (as, for instance, in the fuzzy sphere vacuum where $R=R_{sph}/k$, with $R_{sph}$
the fuzzy sphere radius). As we noted before, there is no $M_P$ inherent to ABJM since pure ABJM is of course conformal but $M_P$ does appear as a 
unit of length in the same way as in AdS/CFT. For example, the physical radius of the spacetime sphere corresponding to the worldvolume
fuzzy sphere is $R^2_{sph}=\frac{2}{N} \Tr\left(X^IX_I^\dagger\right)=2\pi k\mu N l_P^3$. Then $m_s$ and $m_p$, or rather $N_s$ and $N_p$, should be obtained as sizes of the ABJM solutions corresponding to gravitons, 
i.e. sizes of vortices, with $N_s+N_p\leq N$. Unfortunately, for technical reasons we had to restrict ourselves to the case of $N=2$ but we should,
in principle, consider the case $1\ll N_s, N_p\ll N$. It is not clear if it is sufficient to consider $N_s$ vortices on top of each other (i.e. an
charge $N_s$ vortex), or if we need something more sophisticated in the case $1\ll N_s\ll N$. Finally, we should have a corresponding object for $m_1$ and the graviton wavefunction $h_{--}$ should, in principle, match on to a vortex wavefunction. One possibility for the latter would be to define it as $a(r)-n$ or $g(r)-g(\infty)=g(r)-1$, which decays to zero with the mass of the smallest excitation of the theory, i.e. as $e^{-\mu r}$ in the case of the $N_s=1$
vortex. 

\item {\bf Vortex calculations:} The classical scattering of vortices for parallel separation gave 
\be
\frac{1}{k}L_{class}=\sum_i(\mu +\mu \frac{v_i^2}{2})-\sum_{ij} \mu q K_0(2\mu r_{ij})v_{rel,ij}^2\,,
\ee
for $N=2$, i.e. in the $U(2)\times U(2)$ ABJM theory. If we wanted to parallel the computation of BFSS, the next term to calculate would be at one-loop. Specifically, to obtain the interaction potential we must consider the classical solution of two vortices, one at $z=0$ and one at $z^i=b^i+v^it$, and then calculate the one-loop determinants $\sum_n \omega_n /2$ around it. To obtain matching with our tree-level graviton scattering result, we would need at least that the one-loop determinants give a result proportional to $m_sm_pv^4$, but even that seems very difficult to check directly, and we leave it for further work.

In the case of transverse vortex separation, we anticipate the same situation as in BFSS in that,
if we take two vortices in different SU(N) blocks then at the classical level they are non-interacting, even if they have non-vanishing velocities. Even though we will run into the problem of regulating their infinite energies, it is instructive to see how this works in the pure ABJM case. There, for a single vortex, $Y^A=v^A\mathbb{I}$ for $A=2,3,4$. Therefore,
by setting
\be
Y^A(t)=\left(x_1^A+v_1^At\right)\mathbb{I}_{N_1\times N_1}+\left(x_2^A+v_2^At\right)
\mathbb{I}_{N_2\times N_2}\,,
\ee
and the vortices at the same worldvolume positions
in the ABJM action, we obtain the nonrelativistic mass of the supergravitons exactly as in BFSS, 
\be
H\sim \frac{1}{2}\Tr\left|\dot X^i\right|^2+...=E_0+(N_1+N_2) \frac{v^2}{2}\,,
\ee
where $E_0$ is the energy of the free vortices at rest 
and the interaction potential comes from the 1-loop fluctuation around the vortex solution with 
$X^i(t)=b^i+v^it$. There are two immediate subtleties that arise: Firstly, it is not clear whether this will work for both $X^i$ being the four fuzzy sphere directions, as well as for the four remaining transverse directions and, secondly we have no explicit realization of these vortices in the massive ABJM model. Both of these are left for future work.

\item {\bf Gravity calculations:} Even if we could match our graviton scattering result (single graviton exchange)
\be
V_{int}^{(1)}=m_p\frac{v_{rel,sp}^4}{8}h^{++}\propto m_s m_p v_{rel,sp}^2\sim \frac{N_sN_p}{R^2}v_{rel,sp}^4\,,
\ee
with a corresponding vortex computation like the one-loop one suggested above, it remains to find a gravity calculation matching the classical vortex scattering potential. For that, we need a calculation that will give a result proportional 
to $v^2_{rel,ij}$. It does not necessarily need to be a classical calculation but it should certainly be something different from the leading single-graviton exchange calculated above {\it a la} 't Hooft, {\it i.e.} we want to find some term unaccounted for by that calculation. 

One possibility suggests itself immediately. Since the graviton itself should correspond to a vortex, it is not unreasonable to expect that we should also be able to find 
an effective 3-dimensional Lagrangian for {\it interacting supergravitons}. Its construction, however, is plagued with subtleties. The supergraviton (a $D0$-brane) and 
the other $D$-branes of the theory are also solutions of the 10-dimensional supergravity action, but only after introducing explicit delta function source terms in it. 
So we should, in principle, find a solution for interacting supergravitons and plug it back in the action, exactly as we did for the corresponding vortices. But that of course means that we now  also need to do something about the extra dimensions. The most conservative route would be to explicitly perform the integration over the extra dimensions. Finally, one must still subtract the nonzero result for the background without supergravitons, corresponding to the ABJM field theory without vortices. To this end we start with 
\be
{\cal L}_{11}={\cal L}_{EH}+{\cal L}_{matter}+{\cal L}_{source}-({\cal L}_{EH}+{\cal L}_{matter})_{bgr}\,.
\ee
To arrive at the effective Lagrangian, we evaluate ${\cal L}_{11}$ on its supergraviton solution,  and integrate out over all the 
directions (parallel and transverse), obtaining
\footnote{Here $x^-$ is DLCQ time, but we rewrite in terms of usual 
time $t\equiv x_0$ and trivially integrating over $x_{11}$ at fixed 11th momentum of the fields.} 
\be
L_{eff}(\vec{y}_i(t))=
\int d^7\vec{x}d^2\vec{y}\sqrt{g}{\cal L}_{11}(\tilde h_{\mu\nu}(\vec{y}, r; \vec{y}_i(t)))\,,\label{Leff}
\ee
where 
\be
g_{\mu\nu}=g_{\mu\nu}^{(0)}+\tilde h_{\mu\nu}(\vec{x},\vec{y};\vec{y}_i(t))\delta(x_{11}-t)\,, 
\ee
and we drop the $\delta(x_{11}-t)$ factor from the metric together with the integration over $x_{11}$, following \cite{Becker:1997cp}. Note that the $\delta(x^-)$ factor in $h_{--}=\tilde h_{--}\delta(x^-)$ 
can be eliminated from $L_{eff}$ in this way only if it is the same for all the gravitons under consideration. But the different velocities of different gravitons mean that they do not move in exactly the same direction, and the delta function cannot be eliminated! So this approximation is exact only for the first order in the 
expansion in $v$, i.e. for the $v^2$ term we are interested in, but there will be extra terms starting at order $v^4$.
In order to calculate such $v^4$ effects, one must consider the full interacting problem, with differently oriented 
shockwaves. 't Hooft showed that in flat space this is reproduced by the source-probe calculation used by 
\cite{Becker:1997cp}, but now in our curved background the calculation might be incomplete.
Nevertheless, this proposal passes several simple tests: 
\begin{itemize}

\item On the background (in the absence of a supergraviton) 
we have, correctly ${\cal L}_{11}$=0, since ${\cal L}_{source}=0$ by 
definition. 

\item In the presence of a single (free) supergraviton, ${\cal L}_{EH}-({\cal L}_{EH})_{bgr}$ is also zero, since 
the only nonzero equation of motion is $G_{--}-(G_{--})_{bgr}\propto \Delta h_{--}$, and $g^{--}=0$. The only 
contribution to $L_{eff}$ comes from the source term, the integral of ${\cal L}_{source}$, which is exactly 
equal to the energy of the free supergraviton, as it should be.

\item In the case of interacting supergravitons however, we will have a nonzero interaction energy coming from 
${\cal L}_{EH}-({\cal L}_{EH})_{bgr}$. Again, this energy will only be nonzero for a relative velocity. 
If the gravitons are parallel, it will be as if we have a single graviton of source equal to the sum of the all sources, for 
which we get no interaction. Further, Galilean invariance imples that the interaction has to 
enter as integer powers of $(\dot q_i-\dot q_j)^2=v_{rel,ij}^2$, so that we will only get $v^2,v^4,v^6,...$ terms.

\item And finally the 1-loop ABJM interaction appears from classical corrections to the
$\delta(x^-)$-factored Lagrangian as it should (although, of course, there must also be quantum gravity corrections as well, i.e. one 
must consider in the general the quantum effective Lagrangian instead of the classical Lagrangian).
\end{itemize}
\end{itemize}
\subsection{M5 brane backreaction}

As a final point in this section, we now sketch how to construct the solution for adding the backreacted $M5$-brane correction to the background corresponding to massive ABJM. 
As we saw in Section 5, 
in the type IIB theory the background corresponds to a pp-wave, and the $M5$-brane corresponds to a $D3$-brane wrapping the $S^3$ defined by $r_1=$ fixed, 
$r_2=0$ inside the maximally 
supersymmetric pp-wave. Such a solution has been implicitly written in Fig.1f of LLM \cite{Lin:2004nb}. Moreover, if the M-theory dual giant graviton is to have size $R=1/\mu$, so too must the $D3$-giant in IIB. Our strategy will therefore be to {\it(a)} use the LLM prescription to find an explicit solution, {\it (b)} T-dualize this to type IIA and {\it (c)} lift to M-theory.
Recall that the LLM solution for any $\frac{1}{2}$ BPS type IIB supergravity 
\bea
ds^2&=&-h^{-2}(dt+V_idx^i)^2+h^2(dy^2+dx^idx^i)+ye^Gd\Omega_3^2+ye^{-G}d\tilde{\Omega}_3^2\nonumber\\
h^{-2}&=&2y\cosh G\nonumber\\
y\d_y V_i&=&\epsilon_{ij}\d_j z\nonumber\\
y(\d_i V_j-\d_j V_i)&=&\epsilon_{ij}\d_y z\nonumber\\
z&=&\frac{1}{2}\tanh G\\
F&=&dB_t\wedge(dt+V)+B_t dV++d\hat B\nonumber\\
\tilde{F}&=&d\tilde{B}_t\wedge(dt+V)+\tilde{B}_tdV+d\hat{\tilde{B}}\nonumber\\
B_t&=&-\frac{1}{4}y^2e^{2G};\;\;\;\;\;
\tilde{B}_t=-\frac{1}{4}y^2e^{-2G}\nonumber\\
d\hat B &=& -\frac{1}{4}y^3*_{3}d(\frac{z+1/2}{y^2});\;\;\;\;\;
d\hat {\tilde{B}} = -\frac{1}{4}y^3*_{3}d(\frac{z-1/2}{y^2})\nonumber
\eea
is completely determined by a single function $z$ which is itself the solution of the {\it linear} equation
\be
\d_i \d_i z+y\d_y\left(\frac{\d_y z}{y}\right)=0\,.
\ee
and absence of singularities means that $z(x_1,x_2,y=0)=\pm 1/2$, and $z\leftrightarrow -z$ is particle-hole 
duality. Note that the equation for $\Phi=z/y^2$ is the Laplace equation in 6d flat space, with 4d spherical symmetry,
with $y$ the radial coordinate for these 4 dimensions. The solution is 
\bea
z(x_1,x_2,y)&=&-\frac{1}{2\pi}\int_{\d {\cal D}}dl\,\, n_i'\frac{x_i-x_i'}{(\vec{x}-\vec{x}')^2+y^2}+\sigma\nonumber\\[-0.4cm]
&{}&\\
V_i(x_1,x_2,y)&=&\frac{\epsilon_{ij}}{2\pi}\oint_{\d{\cal D}}\frac{dx'_j}{(\vec{x}-\vec{x}')^2+y^2}\nonumber
\eea
where $\sigma$ is a contribution from infinity if $z$ is constant outside of a circle of very large radius 
(asymptotically $AdS_5\times S_5$ geometries), with $\sigma=\pm 1/2$ if $z=\pm 1/2$ asymptotically, ${\cal D}$ 
is a droplet, and $n_i$ is the unit normal vector on it, pointing towards the region of $z=+1/2$.

The solution we want is a superposition of two simple solutions: 

1){\bf The pp-wave}: This solution is obtained by having the boundary condition at $y=0$ as a half-filled plane, {\it i.e.} $z(x_1',x_2',0)=\frac{1}{2} \mathrm{sgn}(x_2')$. Explicitly performing the integrals above gives 
\bea
&& z(x_2,y)=\frac{1}{2}\frac{x_2}{\sqrt{x_2^2+y^2}}\,,\nonumber\\[-0.4cm]
&{}&\\
&& V_1=\frac{1}{2}\frac{1}{\sqrt{x_2^2+y^2}};\;\;\; V_2=0\,.\nonumber
\label{ppsol}
\eea
To put the geometry into the standard pp-wave form
\be
ds^2=-2dtdx_1-(r_1^2+r_2^2)dt^2+d\vec{r}_1^2+d\vec{r}_2^2\,,
\ee
we defined
\be
y=r_1r_2;\;\;\;\;
x_2=\frac{1}{2}(r_1^2-r_2^2)\,.
\label{pptrans}
\ee

2) {\bf The $AdS_5\times S_5$ solution}: 
Now the boundary condition at $y=0$ is a large spherical droplet, i.e. for $\tilde{z}
=z-1/2$, the 6d Laplace equation for $\tilde{\Phi}=\tilde{z}/y^2$ has source on a disk of radius $r_0$. Integrating, one obtains
\bea
&&\tilde{z}(r,y;r_0)=\frac{r^2-r_0^2+y^2}{2\sqrt{(r^2+r_0^2+y^2)-4r^2r_0^2}}-\frac{1}{2}\,,\nonumber\\[-0.4cm]
&{}&\\
&& -r\sin\phi V_1+r\cos\phi V_2=V_\phi(r,y;r_0)=-\frac{1}{2}\left(\frac{r^2+y^2+r_0^2}{\sqrt{(r^2+r_0^2+y^2)^2-4r^2
r_0^2}}-1\right)\,.\nonumber
\label{adssol}
\eea
Defining new coordinates by 
\bea
&&y=r_0\sinh\rho \sin\theta\nonumber\\
&&r=r_0\cosh \rho\cos\theta\\
&&\tilde{\phi}=\phi -t\nonumber
\label{coords}
\eea
puts the metric into the more familiar form,
\be
ds^2=r_0\left[-\cosh^2\rho dt^2 +d\rho^2+\sinh^2\rho d\Omega_3^2+d\theta^2+\cos^2\theta d\tilde{\phi}^2
+\sin^2\theta d\tilde{\Omega}_3^2\right]\,,
\label{adsmetric}
\ee
with the area of the droplet in the $(x_{1},x_{2})$ plane $A=4\pi^2l_P^4$  and where $l_P=g^{1/4}\sqrt{\alpha '}$. This is clearly proportional to $N$, the number of branes. Large $N$ therefore corresponds to a large droplet. That the pp-wave is the Penrose limit of $AdS_5\times S_5$ translates into the fact that near the surface of a large spherical droplet, locally the droplet looks like a half-filled plane.

Since the equation for the only unknown function $z(x_i,y)$ is linear, we can trivially construct a solution 
that on the boundary give superposition of circles (and maybe half-filled planes), by just adding up the solutions 
for $z(x_i,y)$. In particular, for concentric circles we get 
\bea
&&\tilde{z}=\sum_i (-1)^{i+1}\tilde{z}\left(r,y;r_0^{(i)}\right)\,,\nonumber\\[-0.4cm]
&{}&\\
&&V_\phi=\sum_i (-1)^{i+1}V_\phi\left(r,y;r_0^{(i)}\right)\,.\nonumber
\eea
To obtain the solution we want - a $D3$-brane wrapping the $S^3$ with $r_1=r_{1,0}$, $r_2=0$ and $(t,x_1)$ fixed inside the pp-wave - we translate into LLM
language which, from (\ref{pptrans}), means that we need $x_2=\frac{(r_{1,0})^2}{2}$, $y=0$ and $(t,x_1)$ fixed. This is a small circle at a position 
$x_2=\frac{(r_{1,0})^2}{2}$ away from the $x_2=0$ boundary of the half-filled plane. Specifically, we are interested in the 
circle of size $r_{1,0}=1/\mu$ (or, in the notation used here, with $\mu=1$, $r_{1,0}=1$). Thus the solution we seek is the 
(finite energy) solution of Fig.3f in LLM, a distance $x_2=1/2$ from the Fermi surface, and size given by how many branes we want wrapped, in this case one.

In terms of the function $z(x_{1},x_{2},y)$ this is just the sum of the pp-wave 
solution (\ref{ppsol}) and a rescaled version of (\ref{adssol}), of area given by $N=1$ (one M5-brane), where we replace $r$ of (\ref{adssol}) with $|\vec{x}-\vec{x}_0|$, and 
$\vec{x}_0=(0,\frac{1}{2})$. Unfortunately, this solution looks rather complicated, but to obtain the backreacted M5-brane we still need to T-dualize it to type IIA and lift 
up to M-theory. This, we leave as an interesting open problem.

\section{Conclusions}

In this paper we have looked for a Matrix theory-type construction from ABJM model. We have identified a BPS vortex solution of ABJM as corresponding 
to a spacetime supergraviton (a pointlike object carrying $D0$-brane charge). In the case of the pure ABJM model, the identification is problematic since, in addition to the supergraviton, we also have also an infinite energy 2-brane to worry about. We find that this is resolved in the case of the massive maximally supersymmetric deformation of ABJM since, here, the solution 
has finite energy. We have also found a BPS kink solution of ABJM, using a certain limiting procedure. This solution can be identified with 
a $D2$-brane with one direction parallel to the ABJM worldvolume, and one direction transverse. The vacuum supersymmetric fuzzy sphere solution had already been identified with the $D4$-brane wrapped on $S^2$. 

In the latter half of the article, we identified the spacetime background corresponding to the massive ABJM deformation, and computed the leading graviton scattering 
interaction potential using a shockwave-probe method previously developed for the BFSS Matrix theory. We then computed the classical interaction potential of two vortices in 
the $N=2$ ABJM model, and found that it differs from the supergraviton potential calculated by scattering two shockwaves {\it a la} 't Hooft. 
We then speculated on how one could fix this mismatch, noting 
that there are many differences with respect to the BFSS computation. Chief among these is the fact that now, the two spatial worldvolume directions are special. We argued that perhaps one needs to find an effective Lagrangian for supergravitons that would match with the 
effective Lagrangian for vortex interaction, and that our gravity calculation could be matched by a one-loop calculation in the vortex background,
similar to the BFSS case. Ultimately, though, it is clear that much more work is needed to find a concrete Matrix theory construction for ABJM. 

\section*{Acknowledgements}
\noindent 

We would like to thank Toshiaki Fujimori, Shin Sasaki, Muneto Nitta and Giacomo Marmorini for collaboration at the initial stages of this project and Alex Hamilton and Andrea Prinsloo for enjoyable discussion throughout. JM is supported by the National Research Foundation (NRF) of South Africa under the Thuthuka and Key International Scientific Collaboration programs. AM acknowledges support from a grantholder PhD fellowship from the NRF.

\begin{appendix}

\section{Graviton wavefunction equation}

In this Appendix we present some of the details of the solution of the graviton (shockwave)  wavefunction equation, 
\be
\Delta^\perp_{bgr}h_{--}=Q\delta^\perp(\vec{x},\vec{y})\,.
\label{gravitonwave}
\ee
We begin by splitting the problem into two cases depending on whether the 11th dimension is parallel to the worldvolume (transverse speparation) or perpendicular to the worldvolume directions (longitudinal separation).

\begin{itemize}
  \item {\bf Transverse separation:}\\
In this case, $\perp$ denotes directions transverse to the wave. Among the 
$\vec{y}=t,x,y$ directions, however, only $y$ has this property, so eq.(\ref{gravitonwave})
becomes
\be
H^{-1/3}(\vec{x})\left[H(\vec{x})\d_{y}^2+\d^2_{\vec{x}}\right]h_{--}\left(y,\vec{x}\right)
=Q\delta(y)\delta^8(\vec{x})\,.
\ee
After a Fourier transform in the worldvolume direction $y$, 
\be
h_{--}(\vec{x},y)=\int dp\,\, \frac{e^{ipy}}{2\pi}\phi_p(\vec{x})\,,
\ee
we get the equation 
\be
\left(1-\mu^2\vec{x}^2\right)^{-1/3}\left[\d^2_{\vec{x}}-p^2(1-\mu^2\vec{x}^2)\right]\phi_p(\vec{x})=Q\delta^8(\vec{x})\,.
\ee

\item {\bf Parallel separation:}\\
Now, there are two directions, $\vec{y}=(x,y)$ that are transverse to the wave and parallel to the 3-dimensional worldvolume. Expanding the Laplacian and simplifying gives
\be
(1-\mu^2\vec{x}^2)^{-1/3}\left[\d^2_{\vec{x}}-p^2(1-\mu^2\vec{x}^2)\right]\phi_{\vec{p}}(\vec{x})=Q\delta^7(\vec{x})
\ee
where now
\be
h_{--}(\vec{x},\vec{y})=\int d^2\vec{p}\,\, \frac{e^{i\vec{p}\vec{y}}}{2\pi}\phi_{\vec{p}}(\vec{x})\,.
\ee
\end{itemize}
We can actually treat both cases in a unified way since, in either case, we can write the equation that needs to be solved as 
\be
\left[\frac{\d}{\d \vec{x}^2}+\mu^2 p^2 \vec{x}^2-p^2\right]\phi_p(\vec{x})=-Q\delta^d(\vec{x})\,.
\ee
Two points to note about this equation are ({\it a}) that it is manifestly spherically symmetric and ({\it b}) that if we put $Q=0$ and $p=i\tilde{p}$ we get the isotropic harmonic oscillator in $d$ dimensions. To obtain a solution, we need to first impose boundary conditions, either at a finite distance or at infinity. The natural one would be to impose normalizability of the wave function at infinity, but as we will see, this is not a good condition.

To proceed, let's first solve the equation for $Q=0$, using techniques for central interaction potential. Specifically, expand $\phi_{p}$ in spherical harmonics as
\be
\phi_p(\vec{x})=\frac{y_{p,k}(r)}{r^{n/2}}Y_k(\widetilde{\Omega})\,,
\ee
where the $n$-sphere spherical harmonics $Y_{k}^{(n)}$ satisfy, 
\be
\widetilde{\Box}Y_k^{(n)}(\widetilde{\Omega})=-k(k+n-1)Y_k^{(n)}(\widetilde{\Omega})\,.
\ee
Consequently
\be
\left[\frac{d^2}{dr^2}-\frac{k(k+n-1)+n(n-2)/4}{r^2}+\mu^2 p^2 r^2-p^2\right]y_{p,k}(r)=0\,.
\label{yequation}
\ee
Now note that when $p=k=0$ we obtain $r^{n/2}$ and $r^{1-n/2}$ as solutions, since then 
$\phi_p(\vec{x})$ has a constant and $r^{1-n}$ as solutions. Next, on setting $p=i\tilde{p}$, eq.(\ref{yequation}) takes the form
\be
y''-\frac{a}{r^2}y-br^2y+cy=0\,.
\ee
This we recognize as a confluent hypergeometric equation whose solutions are, of course, (a linear combination of) confluent hypergeometric functions $\,{}_{1}F_{1}$ and ${}_{2}F_{1}$,
\be
\!\!\!\!\!y(r)=e^{\frac{-\sqrt{b}r^2}{2}}r^{\frac{1}{2}+\frac{\sqrt{1+4a}}{2}}
\left[C_1\,{}_{1}F_{1}\left(\frac{1}{2}+\frac{\sqrt{1+4a}}{4}-
\frac{c}{4\sqrt{b}},1+\frac{\sqrt{1+4a}}{2}; \sqrt{b}r^2\right)+C_2\,{}_{2}F_{1}(same)\right].
\ee
On substituting $a=\frac{k(k+n-1)+n(n-2)}{4}$, $b=\mu^2\tilde{p}^2$ and $c=\tilde{p}^2$ we get that
\bea
\phi_p(\vec{x})&=&Y_k^{(n)}(\tilde{\Omega})e^{-\mu\tilde{p}r^2/2}r^k\Bigg[C_1\,\,{}_{1}F_{1}\!\!\left(-\frac{\tilde{p}}{4\mu}+\frac{k}{2}
+\frac{n+1}{4}, k+\frac{n+1}{2};\mu \tilde{p}r^2\right)\nonumber\\
&&+C_2\,\,{}_{2}F_{1}\!\!\left(-\frac{\tilde{p}}{4\mu}+\frac{k}{2}
+\frac{n+1}{4}, k+\frac{n+1}{2};\mu \tilde{p}r^2\right)\Bigg]\,.
\eea
Since we have a spherically symmetric source and we are looking for a solution that goes to zero at infinity, we need to choose a spherically symmetric solution, {\it i.e.} $k=0$. Also replacing 
$\tilde{p}=-ip$, we finally get 
\be
\!\!\!\!\!\phi_p(\vec{x})=e^{\frac{i\mu p r^2}{2}}\Bigg[C_1\,{}_{1}F_{1}\left(\frac{ip}{4\mu}+\frac{n+1}{4}, \frac{n+1}{2}; -i\mu p r^2\right)
+C_2\,{}_{2}F_{1}\left(\frac{ip}{4\mu}+\frac{n+1}{4}, \frac{n+1}{2}; 
-i\mu p r^2\right)\Bigg].
\ee
Further, using the relation between ${}_{2}F_{1}$ and ${}_{1}F_{1}$,
\be
 {}_{2}F_{1}(\a,\b;z)=\frac{\Gamma(1-\b)}{\Gamma(1+\a-\b)}\,{}_{1}F_{1}(\a,\b;z)+\frac{\Gamma(1-\b)}{\Gamma(\a)}z^{1-\b}
\,{}_{1}F_{1}(1+\a-\b,2-\b;z)
\ee
we find that near $r=0$ 
\be
\phi_p(\vec{x})=K \frac{e^{\frac{i\mu p r^2}{2}}}{r^{n-1}}\,{}_{1}F_{1}\left(\frac{ip}{4\mu}+\frac{3-n}{4}, 
\frac{3-n}{2};-i\mu p r^2\right)
\ee
with the constant fixed as a function of the source $Q$ (up to now neglected) as follows. Near $r=0$, the full equation, with 
nonzero source $Q$ becomes
\be
\d_{\vec{x}}^2\phi_p(\vec{x})=-Q\delta^{d}(\vec{x})\,,
\ee
which is solved by
\be
\phi_p(\vec{x})=\frac{Q}{(d-2)\Omega_{d-1} r^{d-2}}\,,
\ee
and so the dominating full solution near $r=0$ must be
\be
\phi_p(\vec{x})=\frac{Q}{(n-1)\Omega_n r^{n-1}}e^{i\mu p r^2/2}\,{}_{1}F_{1}\left(\frac{ip}{4\mu}+\frac{3-n}{4}, 
\frac{3-n}{2};-i\mu p r^2\right)\,.
\label{solution}
\ee
Notice that this solution behaves like $1/r^{n-1}$. To this we can add the subleading correction (which behaves like a constant at near zero),
\be
\phi_{p,2}(\vec{x})=\widetilde{K}_p(-i\mu p)^{\frac{n-1}{2}}e^{i\mu p r^2/2}\,{}_{1}F_{1}\left(\frac{ip}{4\mu}+\frac{n+1}{4}, \frac{n+1}{2}; -i\mu p r^2\right)\,,
\label{othersol}
\ee
where $\widetilde{K}_p$ is an arbitrary 
$p-$dependent constant and we have also kept an explicit power of $p$ above in order to emphasize the $z=-i\mu pr^2$ dependence.

Now to impose boundary conditions. We first try normalizability at infinity and find that both solutions behave in the same way as $r\rightarrow\infty$, namely
\bea
\phi_p(\vec{x})&\rightarrow& 
\frac{Q \Gamma\left(\frac{3-n}{2}\right)}{(n-1)\Omega_{n-1}}\Bigg\{ \frac{e^{-\frac{i\mu pr^2}{2}}(-i\mu p r^2)^{\frac{ip}{4\mu}}
(-i\mu p )^{-\frac{3-n}{4}}}{\Gamma\left(\frac{ip}{4\mu}+\frac{3-n}{4}\right)}\nonumber\\
&&+\frac{e^{\frac{i\mu pr^2}{2}}
(i\mu p r^2)^{-\frac{ip}{4\mu}}(i\mu p )^{-\frac{3-n}{4}}}{\Gamma\left(-\frac{ip}{4\mu}+\frac{3-n}{4}\right)}
\Bigg\}\frac{1}{r^{\frac{n+1}{2}}}\propto \frac{1}{r^{\frac{n+1}{2}}}\,,\nonumber\\
\phi_{p,2}(\vec{x})&\rightarrow&
e^{\frac{i\mu p r^2}{2}}\,{}_{1}F_{1}\left(\frac{ip}{4\mu}+\frac{n+1}{4},\frac{n+1}{2};-i\mu p r^2\right)
\sim (...)r^{\frac{ip}{2\mu}-\frac{n+1}{2}}\nonumber\\
&&+(...)e^{\frac{ip\mu r^2}{2}}r^{-\frac{ip}{2\mu}-\frac{n+1}{2}}
\propto \frac{1}{r^{\frac{n+1}{2}}}
\eea
and correctly going to zero at infinity. But this behaviour is non-normalizable, since 
\be
\int d^d x\, \left|\phi_p(\vec{x})\right|^2 \sim \int^R dr \; r^{n}\frac{1}{r^{n+1}}\sim \ln R\,.
\ee
We can extract more information from the solution by taking the $p\rightarrow \infty$ limit of the result and using the fact that the confluent hypergeometric function
\be
\,{}_{1}F_{1}(\a,\b;z)\simeq \frac{\Gamma(\b)}{\sqrt{\pi}}e^{z/2}(-\a z)^{-\frac{\b}{2}+\frac{1}{4}}\cos\left[2\sqrt{-\a z}-\pi(
\frac{\b}{2}-\frac{1}{4})\right]\,,
 \label{largealpha}
\ee
as $|\a|\rightarrow\infty$. This, in turn means that as $p\rightarrow\infty$, 
\be
\phi_p(\vec{x})\simeq \frac{\Gamma\left(\frac{3-n}{2}\right)}{\sqrt{\pi}(n-1)\Omega_{n-1}}
\frac{Q}{r^{n-1}}(\pm ipr)^{\frac{n-2}{2}}\cos\Big[\pm ipr +\pi \frac{n-2}{4}\Big]\,.
\label{resulttwo}
\ee
To continue, we need to fix $\widetilde{K}_p$ by imposing appropriate boundary conditions. Again, we distinguish two cases,

\noindent
{\bf Case 1.} We first assume that we can put $\widetilde{K}_p=0$, thus ignoring $\phi_{p,2}(\vec{x})$. Then, in the {\it transverse separation} case 
where $n=7$, we obtain the graviton wavefunction at $y=0$ by a simple integration of 
$\phi_p(\vec{x})$ over $p$. Assuming also that the large $p$ region dominates the integral, we obtain 
\bea
&&\phi\simeq C\frac{Q}{r^7}\nonumber\\[-0.4cm]
&{}&\\
&&C=\lim_{n\rightarrow 7}\frac{\Gamma\left(\frac{3-n}{2}\right)}{\sqrt{\pi}(n-1)\Omega_{n-1}}(\pm i)^{\frac{n-2}{2}}
\int_{-\infty}^{+\infty} dz z^{\frac{n-2}{2}}\cos\left[\pm iz +\pi \frac{n-2}{4}\right]\,,
\nonumber
\eea
which matches the flat space case (corresponding to the BFSS analysis). Also note that there are no other dimensional parameters
in the $Q/r^7$ behaviour, it is just multiplied by a number (part of the number is an integral over the variable 
$z=pr$), for which we must take a limit since $\Gamma(-2)=\infty$. This independence of the result from $\mu$ is most likely the result of an incorrect initial assumption about the boundary conditions.

In the case of {\it parallel separation}, $n=6$ and the Fourier transform is 
\be
\int \frac{d^2 p}{2\pi}e^{i\vec{p}\vec{y}} \phi_p(\vec{x})
=\int_0^\infty p\; dp \int_0^{2\pi}\frac{d\theta}{2\pi}e^{ipy \cos\theta}\phi_p(\vec{x})\,.
=\int_0^\infty p\; dp \; J_0(py)\phi_p(\vec{x})
\ee
Here, by contrast, we are interested in the $r\rightarrow 0$ limit. If however, $p$ stays finite, we find that the solution (\ref{solution}) becomes $p-$independent, and then the above Fourier transform gives zero (or rather, $\delta(y)$). 
So we must again consider the large $p$ limit in (\ref{resulttwo}). This region of large $p$ can still give a nontrivial contribution to the $p$ integral, since $pr$ is not necessarily small. Then, also substituting $n=6$, gives 
\bea
\phi(y,r\rightarrow 0)&\simeq& -\frac{4Q}{15\Omega_6}\frac{1}{r^3}\int_0^\infty dp\; p^3\; J_0(py)\cos(ipr+ \pi)
=-\frac{4Q}{15\Omega_6}\frac{1}{r^3}\int_0^\infty dp\; p^3\; J_0(py) \cosh (pr)\nonumber\\
&=&-\frac{Q}{15\Omega_6}\frac{1}{r^3}\frac{1}{r^4}{}_2F_1\left(2,5/2;1;-\frac{y^2}{r^2}\right)\,.\label{fresult}
\eea
To understand what the behaviour of the solution near $r=0$ is, the identities  
\bea
\int_0^\infty dx x^{\mu-1}e^{-\a x}J_\nu(\b x)&=&\frac{(\b/(2\a))^\nu\Gamma(\nu+\mu)}{\a^\mu \Gamma(\nu+1)}
{}_2F_1(\frac{\nu+\mu}{2},\frac{\nu+\mu+1}{2};\nu+1;-\frac{\b^2}{\a^2})\nonumber\\
{}_2F_1(\a,\b;\gamma;z)&=&\frac{\Gamma(\gamma)\Gamma(\b-\a)}{\Gamma(\b)\Gamma(\gamma-\a)}(-z)^{-\a}
{}_2F_1(\a,\a+1-\gamma;\a+1-\b;\frac{1}{z})\\
&&+\frac{\Gamma(\gamma)\Gamma(\a-\b)}{\Gamma(\a)\Gamma(\g-\b)}(-z)^{-\b}{}_2F_1(\b,\b+1-\gamma;\b+1-\a;\frac{1}{z})\nonumber
\eea
can be used to show that as $y/r\rightarrow\infty$, the hypergeometric function ${}_2F_1\left(2,5;1;-\frac{y^2}{r^2}\right)\sim -\frac{24}{(y/r)^{10}}$ and, subsequently 
$\displaystyle \phi(y,r\rightarrow 0)\simeq \frac{8Q}{5\Omega_6}\frac{1}{r^7}\frac{1}{(y/r)^{10}}$ near $r=0$. Note then that we can actually set $r=0$ directly and obtain $\phi(r=0)=0$, which is clearly not what we wanted, so again the initial boundary condition was incorrect.

\noindent
{\bf Case 2.} We now look for a more physical boundary condition, and we concentrate on the {\it parallel separation} case. 
In reality, the $r$ space terminates at $r=1/\mu$, so we must impose a boundary 
condition there. This is an apparent singularity that one should be able to continue through, thus the appropriate boundary condition 
at $r=1/\mu$ is Neumann, $\phi_p'(r=1/\mu)=0$. 
The reason is that then the point $r=1/\mu$ acts as the origin in angular coordinates, for which the above condition 
is the only one that makes sense (see for instance \cite{Witten:1998zw}). Thus imposing 
\be
\frac{d}{dr}\phi_p(r)|_{r=1/\mu}=0\,,
\ee
on the function
\be
\phi_p(r)=\frac{Q}{5\Omega_6 r^5}e^{\frac{i\mu pr^2}{2}}\,{}_{1}F_{1}\left(\frac{ip}{4\mu}-\frac{3}{4},-\frac{3}{2};-i\mu p r^2\right)
+\tilde{K}_p(-i\mu p)^{\frac{5}{2}}e^{\frac{i\mu pr^2}{2}}\,{}_{1}F_{1}\left(\frac{ip}{4\mu}+\frac{7}{4};\frac{7}{2};-i\mu pr^2\right)\,,
\label{fullsolution}
\ee
and using the identity
\be
\frac{d}{dz}\,{}_{1}F_{1}(a,b;z)=\frac{a}{b}\,{}_{1}F_{1}(a+1,b+1;z)
\ee
we obtain the condition 
\be
{\frac{5\Omega_6}{Q\mu^5}}\widetilde{K}_p(-i\mu p)^{5/2}=\frac{
\,{}_{1}F_{1}\left(\frac{ip}{4\mu}-\frac{3}{4},-\frac{3}{2};\frac{-ip}{\mu}\right)
\left(1-\frac{5\mu}{ip}\right)+\frac{4}{3}\left(\frac{ip}{4\mu}-\frac{3}{
4}\right)\,{}_{1}F_{1}\left(\frac{ip}{4\mu}+\frac{1}{4},-\frac{1}{2};\frac{-ip}{\mu}\right)}
{-\,{}_{1}F_{1}\left(\frac{ip}{4\mu}+\frac{7}{4},\frac{7}{2};\frac{-i
p}{\mu}\right)+\frac{4}{7}\left(\frac{ip}{4\mu}+\frac{7}{4}\right)\,{}_{1}F_{1}\left(\frac{ip}{4\mu}+\frac{11}{4},\frac{9}{2};\frac{-ip}{\mu}\right)}\,.
\label{ratio}
\ee
We already saw that at $r=0$ 
the Fourier transform of the first term in (\ref{fullsolution}) vanishes, so we need to analyze only the second term. Then 
\be
\phi(y,r=0)=\int\frac{d^2 p}{2\pi}e^{i\vec{p}\cdot\vec{y}} \tilde{K}_p(-i\mu p)^{5/2}
\ee
and the right hand side of (\ref{ratio}) gives in the large $p$ limit
\be
\frac{15}{16}\frac{1+i}{1-i}\left(\frac{ip}{2\mu}\right)^{-5}
\ee
which means that we could
close the contour of integration $\int_{-\infty}^{+\infty} dp$ with a semicircle in the upper half plane
(since $\widetilde{K}_p(-i\mu p)^{5/2}$ goes to zero as $|p|\rightarrow \infty$), and thus if
the integral above would be one-dimensional instead of two, it would be given by the residues at the poles in the upper half $p$ plane. $\widetilde{K}_p$ has poles in the upper half plane of the complex $p$, given by the solutions to the equation
\be
\,{}_{1}F_{1}\left(\frac{ip}{4\mu}+\frac{7}{4},\frac{7}{2};\frac{-i
p}{\mu}\right)=\frac{4}{7}\left(\frac{ip}{4\mu}+\frac{7}{4}\right)\,{}_{1}F_{1}\left(\frac{ip}{4\mu}+\frac{11}{4},\frac{9}{2};\frac{-ip}{\mu}\right).
\ee
We can easily see, however, that $p=0$ is a solution to this equation, so half the residue at $p=0$ will contribute to the integral, giving a constant contribution (independent of $y$) instead of an exponential.

The next solutions that we obtain for the poles (with \texttt{Mathematica}) are $p=0.\pm 9.14066 i\mu$, for which the residue does indeed 
give an exponential. So the main contribution to the 
integral gives a constant, plus an exponential $\sim e^{-m_1y}$.

Here  $m_1$ the lowest imaginary part in the upper half plane among these solutions. Note that we cannot say for
certain that $m_1=9.14066\mu$, since \texttt{Mathematica} searches are always in a neighbourhood; if there is a solution of small imaginary part at large real part, we cannot say from the numerical result. Continuing the numerical search for solutions, $p$, of the root equation, 
we find $0.\pm 13.4767 i\mu; 0.\pm 17.6461 i\mu; 0.\pm 21.7516 i \mu$, with a rather general starting point (even a {\it real} one). This seems to hint that all solutions are pure imaginary? 
If so, we would have a chance to prove that $m_1=9.14066\mu$.  

Thus if the $p$ integral were be one-dimensional, we would get $\phi(y\rightarrow \infty,r=0)\sim e^{-m_1 y}$. 
We can then at least conclude that if we restrict the dependence to only one of the $y$ coordinates, we indeed get 
$\phi(y\rightarrow \infty,r=0)\sim e^{-m_1 y}$. A similar result is expected for a two-dimensional $y$, since then we would need to do the integral
\be
\phi(|y|,r=0)=\int_0^\infty p\, dp\; J_0(py)\tilde{K}_p(-i\mu p)^{5/2}\,,
\ee
which we cannot perform but the small $y$ behaviour would likely be given by expanding 
$\widetilde{K}_p(-i\mu p)^{5/2}$ at large $p$, which gives $\displaystyle \propto \int_0^\infty dp\; p^{-4} J_0(py)\propto y^3$.
It is not very clear how to obtain the large $y$ behaviour, especially since as we saw above, we are interested in a
subleading term. The leading term is indeed (rather easily) obtained as a constant, since  
$|p|\rightarrow 0$, 
\be
\widetilde{K}_p(-i\mu p)^{5/2}\simeq \frac{Q\mu^5}{\Omega_6}\frac{63}{2(p/\mu)^2}\,,
\ee 
and
\be
\phi(|y|,r=0)\propto\int_0^\infty \frac{dp}{p}J_0(py)={\rm const}.
\ee
(the exact form of the constant we cannot be sure of, since the integral above is outside the range of validity
for formulas we could find). 
It is very likely then that the same massive behaviour $e^{-m_1y}$ persists for the first subleading term but this needs to be checked.
\end{appendix}

\bibliographystyle{utphys}
\bibliography{biblioabjm}

\providecommand{\href}[2]{#2}\begingroup\raggedright\begin{thebibliography}{10}

\bibitem{Bagger:2006sk}
J.~Bagger and N.~Lambert, ``{Modeling multiple M2's},''
  \href{http://dx.doi.org/10.1103/PhysRevD.75.045020}{{\em Phys. Rev.} {\bf
  D75} (2007)  045020},
\href{http://arxiv.org/abs/hep-th/0611108}{{\tt arXiv:hep-th/0611108}}.

\bibitem{Bagger:2007vi}
J.~Bagger and N.~Lambert, ``{Comments On Multiple M2-branes},''
  \href{http://dx.doi.org/10.1088/1126-6708/2008/02/105}{{\em JHEP} {\bf 02}
  (2008)  105},
\href{http://arxiv.org/abs/0712.3738}{{\tt arXiv:0712.3738 [hep-th]}}.

\bibitem{Bagger:2007jr}
J.~Bagger and N.~Lambert, ``{Gauge Symmetry and Supersymmetry of Multiple
  M2-Branes},'' \href{http://dx.doi.org/10.1103/PhysRevD.77.065008}{{\em Phys.
  Rev.} {\bf D77} (2008)  065008},
\href{http://arxiv.org/abs/0711.0955}{{\tt arXiv:0711.0955 [hep-th]}}.

\bibitem{Gustavsson:2007vu}
A.~Gustavsson, ``{Algebraic structures on parallel M2-branes},''
  \href{http://dx.doi.org/10.1016/j.nuclphysb.2008.11.014}{{\em Nucl. Phys.}
  {\bf B811} (2009)  66--76},
\href{http://arxiv.org/abs/0709.1260}{{\tt arXiv:0709.1260 [hep-th]}}.

\bibitem{Aharony:2008ug}
O.~Aharony, O.~Bergman, D.~L. Jafferis, and J.~Maldacena, ``{N=6 superconformal
  Chern-Simons-matter theories, M2-branes and their gravity duals},''
  \href{http://dx.doi.org/10.1088/1126-6708/2008/10/091}{{\em JHEP} {\bf 10}
  (2008)  091},
\href{http://arxiv.org/abs/0806.1218}{{\tt arXiv:0806.1218 [hep-th]}}.

\bibitem{Nastase:2009ny}
H.~Nastase, C.~Papageorgakis, and S.~Ramgoolam, ``{The fuzzy $S^2$ structure of
  M2-M5 systems in ABJM membrane theories},''
  \href{http://dx.doi.org/10.1088/1126-6708/2009/05/123}{{\em JHEP} {\bf 05}
  (2009)  123},
\href{http://arxiv.org/abs/0903.3966}{{\tt arXiv:0903.3966 [hep-th]}}.

\bibitem{Nastase:2009zu}
H.~Nastase and C.~Papageorgakis, ``{Fuzzy Killing Spinors and Supersymmetric D4
  action on the Fuzzy 2-sphere from the ABJM Model},''
  \href{http://dx.doi.org/10.1088/1126-6708/2009/12/049}{{\em JHEP} {\bf 12}
  (2009)  049},
\href{http://arxiv.org/abs/0908.3263}{{\tt arXiv:0908.3263 [hep-th]}}.

\bibitem{Banks:1996vh}
T.~Banks, W.~Fischler, S.~H. Shenker, and L.~Susskind, ``{M theory as a matrix
  model: A conjecture},''
  \href{http://dx.doi.org/10.1103/PhysRevD.55.5112}{{\em Phys. Rev.} {\bf D55}
  (1997)  5112--5128},
\href{http://arxiv.org/abs/hep-th/9610043}{{\tt arXiv:hep-th/9610043}}.

\bibitem{Castelino:1997rv}
J.~Castelino, S.~Lee, and W.~Taylor, ``{Longitudinal 5-branes as 4-spheres in
  matrix theory},'' \href{http://dx.doi.org/10.1016/S0550-3213(98)00291-0}{{\em
  Nucl. Phys.} {\bf B526} (1998)  334--350},
\href{http://arxiv.org/abs/hep-th/9712105}{{\tt arXiv:hep-th/9712105}}.

\bibitem{Grosse:1996mz}
H.~Grosse, C.~Klimcik, and P.~Presnajder, ``{Finite quantum field theory in
  noncommutative geometry},'' \href{http://dx.doi.org/10.1007/BF02099720}{{\em
  Commun. Math. Phys.} {\bf 180} (1996)  429--438},
\href{http://arxiv.org/abs/hep-th/9602115}{{\tt arXiv:hep-th/9602115}}.

\bibitem{Kabat:1997im}
D.~N. Kabat and W.~Taylor, ``{Spherical membranes in matrix theory},'' {\em
  Adv. Theor. Math. Phys.} {\bf 2} (1998)  181--206,
\href{http://arxiv.org/abs/hep-th/9711078}{{\tt arXiv:hep-th/9711078}}.

\bibitem{Gomis:2008vc}
J.~Gomis, D.~Rodriguez-Gomez, M.~Van~Raamsdonk, and H.~Verlinde, ``{A Massive
  Study of M2-brane Proposals},''
  \href{http://dx.doi.org/10.1088/1126-6708/2008/09/113}{{\em JHEP} {\bf 09}
  (2008)  113},
\href{http://arxiv.org/abs/0807.1074}{{\tt arXiv:0807.1074 [hep-th]}}.

\bibitem{Sen:1997we}
A.~Sen, ``{D0 branes on T(n) and matrix theory},'' {\em Adv. Theor. Math.
  Phys.} {\bf 2} (1998)  51--59,
\href{http://arxiv.org/abs/hep-th/9709220}{{\tt arXiv:hep-th/9709220}}.

\bibitem{Seiberg:1997ad}
N.~Seiberg, ``{Why is the matrix model correct?},''
  \href{http://dx.doi.org/10.1103/PhysRevLett.79.3577}{{\em Phys. Rev. Lett.}
  {\bf 79} (1997)  3577--3580},
\href{http://arxiv.org/abs/hep-th/9710009}{{\tt arXiv:hep-th/9710009}}.

\bibitem{Becker:1997cp}
K.~Becker and M.~Becker, ``{On graviton scattering amplitudes in M-theory},''
  \href{http://dx.doi.org/10.1103/PhysRevD.57.6464}{{\em Phys. Rev.} {\bf D57}
  (1998)  6464--6470},
\href{http://arxiv.org/abs/hep-th/9712238}{{\tt arXiv:hep-th/9712238}}.

\bibitem{'tHooft:1987rb}
G.~'t~Hooft, ``{Graviton Dominance in Ultrahigh-Energy Scattering},''
\href{http://dx.doi.org/10.1016/0370-2693(87)90159-6}{{\em Phys. Lett.} {\bf
  B198} (1987)  61--63}.

\bibitem{Kang:2004yk}
K.~Kang and H.~Nastase, ``{Planckian scattering effects and black hole
  production in low M(Pl) scenarios},''
  \href{http://dx.doi.org/10.1103/PhysRevD.71.124035}{{\em Phys. Rev.} {\bf
  D71} (2005)  124035},
\href{http://arxiv.org/abs/hep-th/0409099}{{\tt arXiv:hep-th/0409099}}.

\bibitem{Kang:2004jd}
K.~Kang and H.~Nastase, ``{High energy QCD from Planckian scattering in AdS and
  the Froissart bound},''
  \href{http://dx.doi.org/10.1103/PhysRevD.72.106003}{{\em Phys. Rev.} {\bf
  D72} (2005)  106003},
\href{http://arxiv.org/abs/hep-th/0410173}{{\tt arXiv:hep-th/0410173}}.

\bibitem{Nastase:1998sy}
H.~Nastase, M.~A. Stephanov, P.~van Nieuwenhuizen, and A.~Rebhan,
  ``{Topological boundary conditions, the BPS bound, and elimination of
  ambiguities in the quantum mass of solitons},''
  \href{http://dx.doi.org/10.1016/S0550-3213(98)00773-1}{{\em Nucl. Phys.} {\bf
  B542} (1999)  471--514},
\href{http://arxiv.org/abs/hep-th/9802074}{{\tt arXiv:hep-th/9802074}}.

\bibitem{Berenstein:2002jq}
D.~E. Berenstein, J.~M. Maldacena, and H.~S. Nastase, ``{Strings in flat space
  and pp waves from N = 4 super Yang Mills},'' {\em JHEP} {\bf 04} (2002)  013,
\href{http://arxiv.org/abs/hep-th/0202021}{{\tt arXiv:hep-th/0202021}}.

\bibitem{Lowe:2003qy}
D.~A. Lowe, H.~Nastase, and S.~Ramgoolam, ``{Massive IIA string theory and
  matrix theory compactification},''
  \href{http://dx.doi.org/10.1016/S0550-3213(03)00547-9}{{\em Nucl. Phys.} {\bf
  B667} (2003)  55--89},
\href{http://arxiv.org/abs/hep-th/0303173}{{\tt arXiv:hep-th/0303173}}.

\bibitem{Gomis:2008be}
J.~Gomis, D.~Rodriguez-Gomez, M.~Van~Raamsdonk, and H.~Verlinde,
  ``{Supersymmetric Yang-Mills Theory From Lorentzian Three- Algebras},''
  \href{http://dx.doi.org/10.1088/1126-6708/2008/08/094}{{\em JHEP} {\bf 08}
  (2008)  094},
\href{http://arxiv.org/abs/0806.0738}{{\tt arXiv:0806.0738 [hep-th]}}.

\bibitem{Mukhi:2008ux}
S.~Mukhi and C.~Papageorgakis, ``{M2 to D2},''
  \href{http://dx.doi.org/10.1088/1126-6708/2008/05/085}{{\em JHEP} {\bf 05}
  (2008)  085},
\href{http://arxiv.org/abs/0803.3218}{{\tt arXiv:0803.3218 [hep-th]}}.

\bibitem{Arai:2008kv}
M.~Arai, C.~Montonen, and S.~Sasaki, ``{Vortices, Q-balls and Domain Walls on
  Dielectric M2- branes},''
\href{http://arxiv.org/abs/0812.4437}{{\tt arXiv:0812.4437 [hep-th]}}.

\bibitem{Kim:2009ny}
C.~Kim, Y.~Kim, O.-K. Kwon, and H.~Nakajima, ``{Vortex-type Half-BPS Solitons
  in ABJM Theory},'' \href{http://dx.doi.org/10.1103/PhysRevD.80.045013}{{\em
  Phys. Rev.} {\bf D80} (2009)  045013},
\href{http://arxiv.org/abs/0905.1759}{{\tt arXiv:0905.1759 [hep-th]}}.

\bibitem{Auzzi:2009es}
R.~Auzzi and S.~Prem~Kumar, ``{Non-Abelian Vortices at Weak and Strong Coupling
  in Mass Deformed ABJM Theory},''
\href{http://arxiv.org/abs/0906.2366}{{\tt arXiv:0906.2366 [hep-th]}}.

\bibitem{Callan:1997kz}
C.~G. Callan and J.~M. Maldacena, ``{Brane dynamics from the Born-Infeld
  action},'' \href{http://dx.doi.org/10.1016/S0550-3213(97)00700-1}{{\em Nucl.
  Phys.} {\bf B513} (1998)  198--212},
\href{http://arxiv.org/abs/hep-th/9708147}{{\tt arXiv:hep-th/9708147}}.

\bibitem{Cook:2003rx}
P.~L.~H. Cook, R.~de~Mello~Koch, and J.~Murugan, ``{Non-Abelian BIonic brane
  intersections},'' \href{http://dx.doi.org/10.1103/PhysRevD.68.126007}{{\em
  Phys. Rev.} {\bf D68} (2003)  126007},
\href{http://arxiv.org/abs/hep-th/0306250}{{\tt arXiv:hep-th/0306250}}.

\bibitem{Gibbons:1997xz}
G.~W. Gibbons, ``{Born-Infeld particles and Dirichlet p-branes},''
  \href{http://dx.doi.org/10.1016/S0550-3213(97)00795-5}{{\em Nucl. Phys.} {\bf
  B514} (1998)  603--639},
\href{http://arxiv.org/abs/hep-th/9709027}{{\tt arXiv:hep-th/9709027}}.

\bibitem{Gauntlett:1997ss}
J.~P. Gauntlett, J.~Gomis, and P.~K. Townsend, ``{BPS bounds for worldvolume
  branes},'' {\em JHEP} {\bf 01} (1998)  003,
\href{http://arxiv.org/abs/hep-th/9711205}{{\tt arXiv:hep-th/9711205}}.

\bibitem{FIKS}
T.~Fujimori, K.~Iwasaki, Y.~Kobayashi, and S.~Sasaki,
``in preparation,''.

\bibitem{Gomis:2008cv}
J.~Gomis, A.~J. Salim, and F.~Passerini, ``{Matrix Theory of Type IIB Plane
  Wave from Membranes},''
  \href{http://dx.doi.org/10.1088/1126-6708/2008/08/002}{{\em JHEP} {\bf 08}
  (2008)  002},
\href{http://arxiv.org/abs/0804.2186}{{\tt arXiv:0804.2186 [hep-th]}}.

\bibitem{Hosomichi:2008jb}
K.~Hosomichi, K.-M. Lee, S.~Lee, S.~Lee, and J.~Park, ``{N=5,6 Superconformal
  Chern-Simons Theories and M2-branes on Orbifolds},''
  \href{http://dx.doi.org/10.1088/1126-6708/2008/09/002}{{\em JHEP} {\bf 09}
  (2008)  002},
\href{http://arxiv.org/abs/0806.4977}{{\tt arXiv:0806.4977 [hep-th]}}.

\bibitem{Buscher:1985kb}
T.~H. Buscher, ``{Quantum corrections and extended supersymmetry in new sigma
  models},''
\href{http://dx.doi.org/10.1016/0370-2693(85)90870-6}{{\em Phys. Lett.} {\bf
  B159} (1985)  127}.

\bibitem{Buscher:1987sk}
T.~H. Buscher, ``{A Symmetry of the String Background Field Equations},''
\href{http://dx.doi.org/10.1016/0370-2693(87)90769-6}{{\em Phys. Lett.} {\bf
  B194} (1987)  59}.

\bibitem{Buscher:1987qj}
T.~H. Buscher, ``{Path Integral Derivation of Quantum Duality in Nonlinear
  Sigma Models},''
\href{http://dx.doi.org/10.1016/0370-2693(88)90602-8}{{\em Phys. Lett.} {\bf
  B201} (1988)  466}.

\bibitem{Ramgoolam:2001zx}
S.~Ramgoolam, ``{On spherical harmonics for fuzzy spheres in diverse
  dimensions},'' \href{http://dx.doi.org/10.1016/S0550-3213(01)00315-7}{{\em
  Nucl. Phys.} {\bf B610} (2001)  461--488},
\href{http://arxiv.org/abs/hep-th/0105006}{{\tt arXiv:hep-th/0105006}}.

\bibitem{Janssen:2004jz}
B.~Janssen, Y.~Lozano, and D.~Rodriguez-Gomez, ``{Giant gravitons in AdS(3) x
  S**3 x T**4 as fuzzy cylinders},''
  \href{http://dx.doi.org/10.1016/j.nuclphysb.2005.01.022}{{\em Nucl. Phys.}
  {\bf B711} (2005)  392--406},
\href{http://arxiv.org/abs/hep-th/0406148}{{\tt arXiv:hep-th/0406148}}.

\bibitem{Hanaki:2008cu}
K.~Hanaki and H.~Lin, ``{M2-M5 Systems in N=6 Chern-Simons Theory},''
  \href{http://dx.doi.org/10.1088/1126-6708/2008/09/067}{{\em JHEP} {\bf 09}
  (2008)  067},
\href{http://arxiv.org/abs/0807.2074}{{\tt arXiv:0807.2074 [hep-th]}}.

\bibitem{Jevicki:2005ms}
A.~Jevicki and H.~Nastase, ``{Towards S matrices on flat space and pp waves
  from SYM},''
\href{http://arxiv.org/abs/hep-th/0501013}{{\tt arXiv:hep-th/0501013}}.

\bibitem{Becker:1997xw}
K.~Becker, M.~Becker, J.~Polchinski, and A.~A. Tseytlin, ``{Higher order
  graviton scattering in M(atrix) theory},''
  \href{http://dx.doi.org/10.1103/PhysRevD.56.R3174}{{\em Phys. Rev.} {\bf D56}
  (1997)  3174--3178},
\href{http://arxiv.org/abs/hep-th/9706072}{{\tt arXiv:hep-th/9706072}}.

\bibitem{Lin:2004nb}
H.~Lin, O.~Lunin, and J.~M. Maldacena, ``{Bubbling AdS space and 1/2 BPS
  geometries},'' {\em JHEP} {\bf 10} (2004)  025,
\href{http://arxiv.org/abs/hep-th/0409174}{{\tt arXiv:hep-th/0409174}}.

\bibitem{Witten:1998zw}
E.~Witten, ``{Anti-de Sitter space, thermal phase transition, and confinement
  in gauge theories},'' {\em Adv. Theor. Math. Phys.} {\bf 2} (1998)  505--532,
\href{http://arxiv.org/abs/hep-th/9803131}{{\tt arXiv:hep-th/9803131}}.

\end{thebibliography}\endgroup

\end{document}